\documentclass[preprint,12pt]{elsarticle}
\usepackage{amssymb}
\usepackage{amsmath}
\usepackage{makecell} 
\usepackage{booktabs} 

\journal{Pattern Recognition}

\begin{document}
\begin{frontmatter}



\title{Deep Convolutional Neural Networks for Conditioning Extremely Noisy Signals}


\author[inst1]{Andrea Faúndez-Quezada}

\affiliation[inst1]{organization={Optics and Photonics Research Group, Faculty of Engineering},
            addressline={University of Nottingham}, 
            city={Nottingham},
            postcode={NG7 2RD}, 
            state={Nottinghamshire},
            country={United Kingdom}}

\author[inst1]{Salvatore La Cavera III}
\author[inst2]{Giovanna Martínez-Arellano}

\author[inst1]{Sidahmed A Abayzeed*}

\affiliation[inst2]{organization={Institute for Advanced Manufacturing, University of Nottingham},
            addressline={Nottingham}, 
            postcode={NG8 1BB}, 
            state={Nottinghamshire},
            country={United Kingdom}}

\begin{abstract}
We present a deep-learning approach for extracting low-voltage signals occurring over a bandwidth of 10 kHz in the presence of substantial measurement noise. This work is motivated by the need to augment label-free microscopy measurements of electrical signalling in living cells, which suffer from a low signal-to-noise ratio (SNR). We investigated a two-step process involving signal detection using a ResNet classifier, followed by signal extraction with several convolutional neural network (CNN) architectures, including a U-Net and a modified Multilevel Wavelet Convolutional Neural Network (MWCNN). To evaluate these architectures, synthetic data were generated using voltage pulse trains with varying periods, duty cycles, and amplitudes, corrupted with different levels of Gaussian noise to simulate measurement conditions with SNR as low as -20 dB. 

The ResNet classifier successfully discriminated signals from pure noise, achieving an accuracy of over 96\% for SNR greater than -12.5 dB. Accuracy decreased to 92\%, 81\%, and 66\% for SNR of -15 dB, -17.5 dB, and -20 dB, respectively. The target signal can be successfully estimated with an average SNR improvement of 26 dB for input SNR ranging from -5 to -20 dB using the modified MWCNN. However, performance decreases as the input SNR decreases, with an average SNR improvement of 27.9 dB for an input SNR of -5 dB compared to 24.5 dB for -20 dB. These findings suggest that CNN architectures can effectively estimate fast signals over a 10 kHz bandwidth and SNR as low as -20 dB, offering potential applications across a range of measurement fields, including electrophysiology, optics, and communications.
\end{abstract}

\begin{graphicalabstract}
\includegraphics[width=1.05\textwidth]{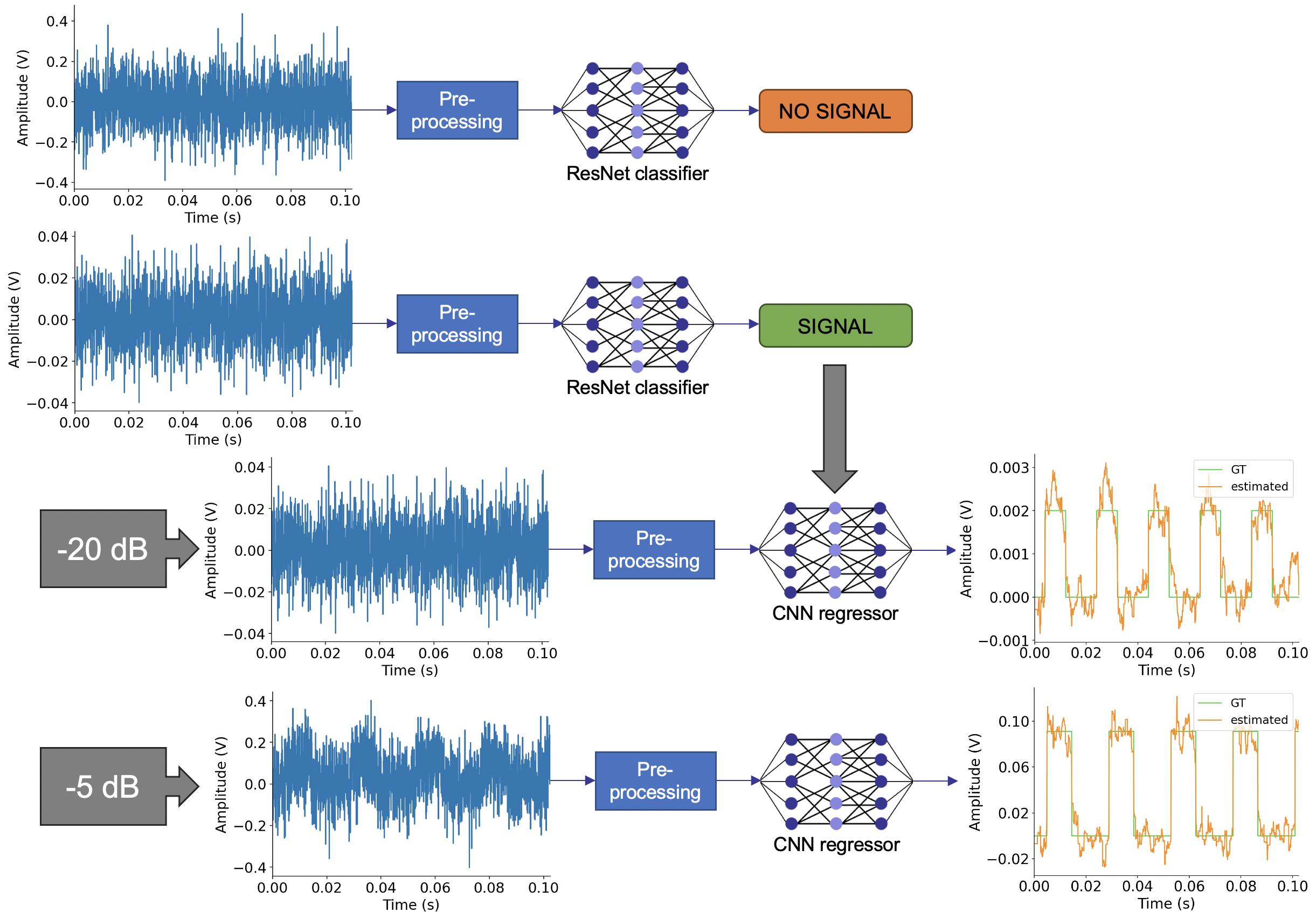}
\end{graphicalabstract}

\begin{highlights}
\item Deep learning reconstructs 10 kHz low-voltage signals with SNR as low as -20 dB.
\item ResNet classifier detects signals with 96\% accuracy for input SNR greater than -12.5 dB.
\item Modified MWCNN improves SNR by up to 27.9 dB in noisy conditions.

\end{highlights}


\end{frontmatter}

\section{Introduction}
\label{introduction}
Living biological cells communicate via electrical signals to maintain key physiological processes such as information processing or heart function \cite{singh2012bioelectrical}. Research has shown that bioelectrical signalling plays a wider role, for instance, in development and morphogenesis, with alteration in diseases such as cancer \cite{levin2021bioelectric}. Measuring these signals generates a wealth of information about human health. Advancements in physics, electronics, and computer science, including sensors and signal processing, have significantly impacted the biological and biomedical domains, revolutionising clinical data acquisition and diagnosis \cite{martinek2021advanced}. Key signals include electromyograms (EMG), electroencephalograms (EEG), and electrocardiograms (ECG), representing the collective bioelectrical activity of skeletal muscle, brain regions, and cardiac cells \cite{martinek2021advanced}. On the other hand, significant research has been directed at understanding bioelectrical signals at the cellular and sub-cellular levels. This is currently studied by measuring membrane potential using intracellular techniques such as patch clamp electrophysiology or micro-electrode methods, which can cause cell damage \cite{xu2021cell}. Furthermore, voltage-sensitive fluorescent techniques are commonly used, but they have limited measurement duration due to photobleaching and toxicity \cite{xu2021cell}.

More recently, several methods were introduced for label-free extended measurements of bio-electrical signals including surface plasmonics \cite{habib2019electro}, quantum sensors \cite{barry2016optical}, graphene electric field sensors \cite{balch2021graphene}. For instance, Surface Plasmon Resonance (SPR) sensors provide a new method for sensing voltage \cite{abayzeed2017sensitive} and microscopic electrical impedance \cite{abayzeed2020plasmonic}, which has been applied to measure membrane potential dynamics from single neurons  \cite{liu2017plasmonic} and cardiomyocytes \cite{habib2019electro}. This non-invasive and label-free approach enables the study of cellular electrical activity without invasive micro-electrodes or fluorescent dyes that can affect cell viability \cite{kim2008optical, chieng2019recent}. However, the prevailing challenge of these techniques is the limited signal-to-noise ratio (SNR) of the acquired data \cite{abayzeed2017sensitive, kim2012vivo}. 

To address the challenge of low SNR in new voltage microscopy techniques, machine learning approaches represent a promising avenue, benefiting from their wide use in denoising biomedical signals. This new approach has provided the ability to capture complex signal patterns while preserving signal features with adaptability to varying noise patterns \cite{rasti2022deep, lin2023ecg, noiseChiang}. In particular, CNNs are capable of learning hierarchical representations of signals such as ECG \cite{rasti2022deep}. This feature is beneficial for denoising bioelectrical signals as it enables the network to capture both low-level features, such as individual waveforms, and high-level features, such as temporal patterns. Recent works have achieved promising results by applying CNNs to improve the SNR of ECG signals. For instance, 2D CNNs were applied to denoise ECG signals \cite{rasti2022deep}, where the data were pre-processed by dividing the ECG signal into segments based on each R-peak. Subsequently, the segments were stacked according to the R-peak locations, resulting in a 2D stacked cardiac cycle tensor that served as input to a SCEN-Net where convolutional blocks were designed to provide encoder-decoder architecture which is specialised in image denoising. Lastly, the output data is transformed to the 1D original dimension of ECG. Furthermore, studies employed CNNs and Long Short-Term Memory (LSTM) models to denoise ECG signals, outperforming traditional methods including wavelets \cite{DLECG}. These models demonstrated superior performance and shorter training times, offering new avenues for bioelectrical signal denoising.

In the present study, we explore deep-learning approaches, with a particular focus on CNNs, to detect and extract signals in extremely noisy conditions. The work is motivated by enhancing SNR in label-free optical imaging of bio-electrical signals. Denoising bioelectrical signals, such as action potential is challenging since the measurements are performed with a bandwidth of 10 kHz \cite{schaefer2006neuronal}. Recent research has investigated deep learning approaches to address the challenges of low SNR (\cite{lacy2024machine,mvuh2024multichannel}; however, our work investigated the extraction of signals with a bandwidth of 10 kHz from extremely noisy measurement, for the first time, to the best of our knowledge. 

To address this challenge, we generated a pulse train with varied amplitude and signal duration as a target signal which was later corrupted by Gaussian noise producing various SNR levels down to -20 dB. This is based on synthetic data which is often used for training neural networks and has previously shown promising results \cite{rasti2022deep, DLECG, martinek2021advanced, antczak2018deep}. Our study has tested four different CNN architectures for uncovering target signals at SNR ranging from -5 to -20 dB. The models demonstrated excellent performance with output SNR ranging from 4.53 $\pm$ 2.55 dB to 22.86 $\pm$ 3.08 dB. The work reported has followed a robust approach where the signal parameters such as amplitude, period and duty cycles were varied providing short-lived signals  with a rising edge of 50 $\mu$s that are extremely challenging to denoise.
In the following sections, we describe the methods and discuss the key results of the study.

\section{Methods}
\label{methods}
\begin{figure*}[htbp]
    \centering
    \includegraphics[width=0.8\linewidth]{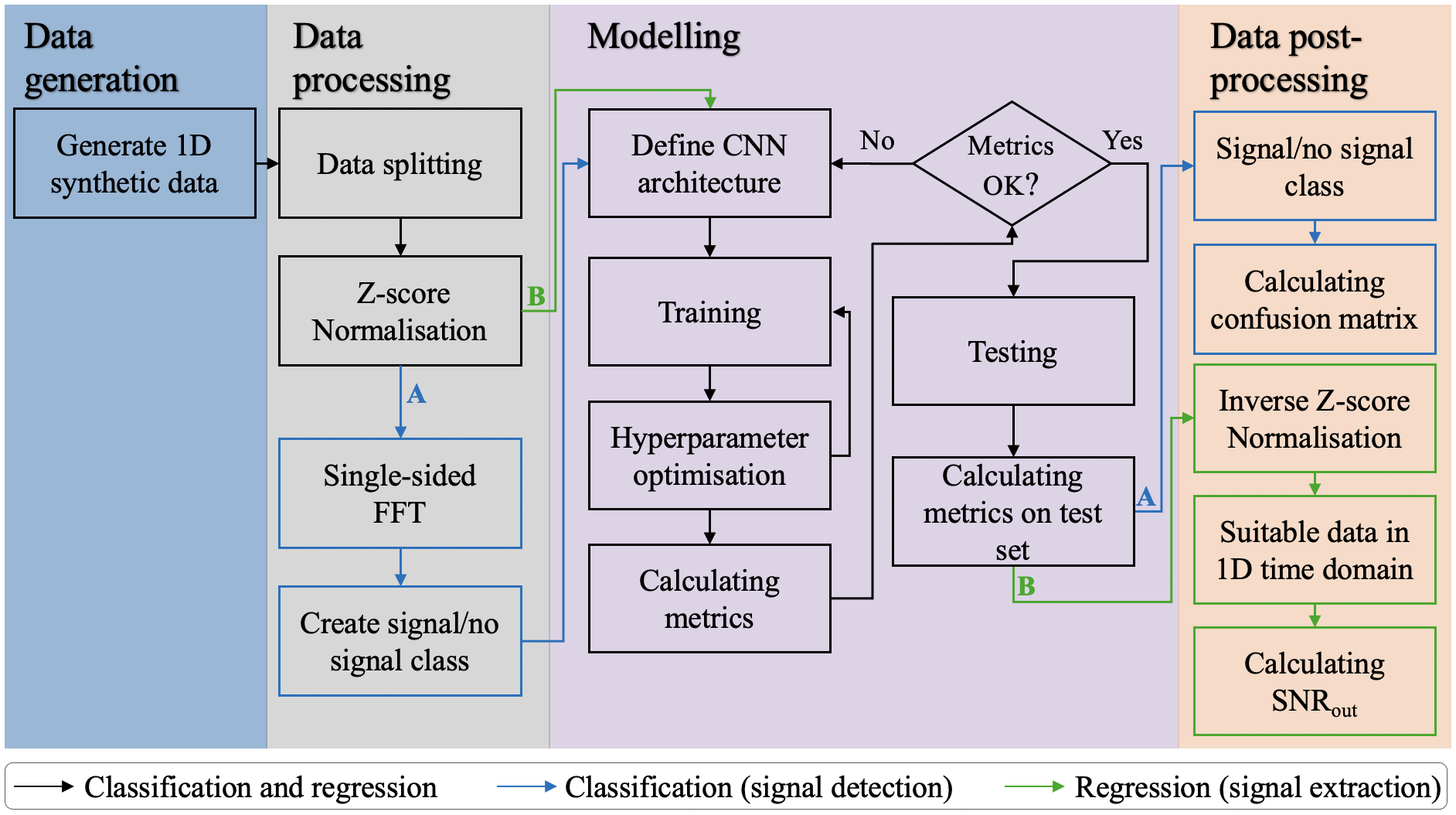}
    \caption[Deep CNNs-based approach for detection and extraction of signals in low SNR conditions.]{Deep CNNs based approach for detection and extraction of signals in low SNR conditions. It starts with data generation, followed by pre-processing that involves two distinct paths: path (A) for signal detection  and path (B) for signal denoising. Next, the modelling stage focuses on training various CNNs for classification and regression. The hyperparameters were adjusted iteratively during this stage based on the metrics obtained from the validation set, continuing until the loss exhibited a consistent plateau for five consecutive epochs. Finally, the post-processing branches into two outcomes: a signal or no signal outputs for the detection, and a denoised estimated signal that is then denormalised.}
    \label{fig:methodology}
\end{figure*}

This section presents the steps employed to detect and extract signals using deep convolutional neural networks for classification and regression. These tasks comprised the following stages: data generation, data processing, modelling, and post-processing, as illustrated in Fig.\ref{fig:methodology}. The approach consisted of two paths (A and B) for classification and regression respectively. In practical situations, the classification could proceed and inform the signal estimation using CNN-based regression.

The proposed methodology is in line with previous machine learning research  \cite{rasti2022deep, DLECG, liu2019multi, DDEECG}, and follows a structure similar to the CRISP-DM methodology \cite{schroer2021systematic}. The stages are detailed in the following subsections and were performed using Python 3.10.12 for data processing, and Tensorflow framework version 2.15.0 for data modelling. The models were executed on a workstation equipped with an Intel(R) Core(TM) i9-10900KF CPU (10 cores, 3.70GHz), 128GB of RAM, and an NVIDIA GeForce RTX 3090 GPU. The system ran on Ubuntu 22.04.1 LTS with 20 concurrent sibling processes.

\subsection{Data generation}
The first stage is data generation, crucial for obtaining the dataset used in training, validation, and testing. Voltage pulse trains with a rising and falling edge of 50 $\mu$s were selected as a model signal to demonstrate the signal detection and extraction. The voltage pulse trains were generated for different amplitudes varying between 1 mV and 150 mV with 1 mV increments. Furthermore, the period was varied from 5 to 25 milliseconds with 1 ms increments, including both values. Similarly, the duty cycle ranged from 0.1 to 0.5 with 0.1 steps. These parameters mirror the action potential, characterised by a sharp peak due the depolarisation and repolarisation phases of the cell membrane, lasting from 1 to 2 milliseconds. This is followed by a refractory period of about 1 to 2 milliseconds, depending on the axon diameter, before another repolarisation occurs \cite{tortora}, yielding a maximum duty cycle of 0.5. Finally, discrete SNR values to generate the noise were set from -5 to -20 dB with step 5, generating 4 parameters. This gives a total of 63,000 unique combinations of parameters (150x21x5x4) originating 15,750 unique ground truth signals, and 63,000 unique noisy signals. The number of samples (i.e., data points) per signal was set at 2,048 for a duration of 0.1024 seconds, allowing for the observation of 4 to 21 pulses with a 20 kHz sampling frequency.

\begin{figure}[ht!]
    \centering
    \includegraphics[width=0.85\linewidth]{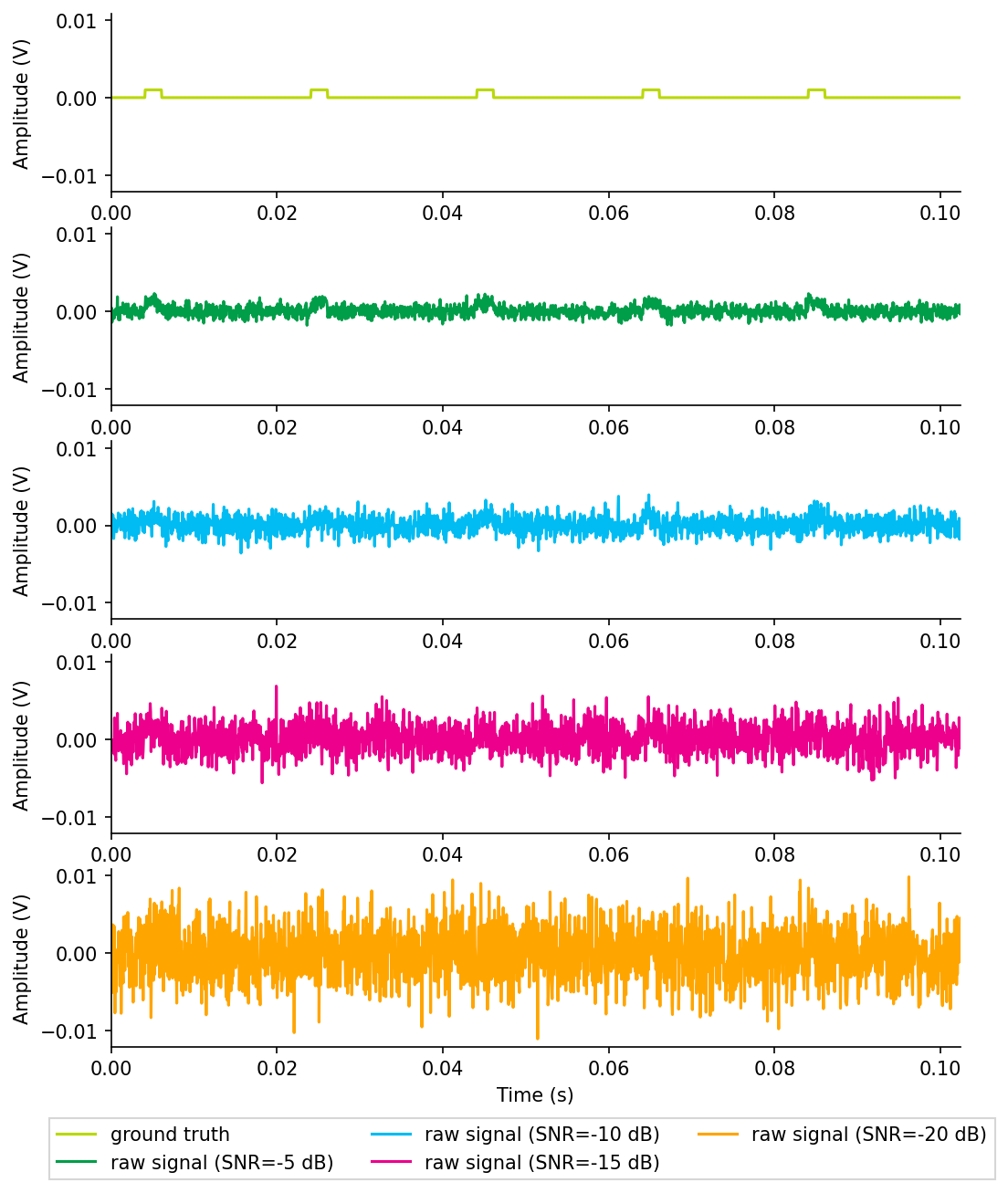}
    \caption[Example of a created signal.]{Example of a synthetic signal of period 20 ms, amplitude 1 mV and duty cycle of 0.1 shown in lime green. The corrupted signals with SNR of -5 dB, -10 dB, -15 dB, and -20 dB are represented in green, blue, pink and orange, respectively.}
    \label{fig:examplesignal}
\end{figure}

To generate noisy signals with specific SNR values, the power of the ground truth signal was first computed, and then the noise power was calculated by using
\begin{equation}
SNR_{dB} = 10 \log_{10}\left(\frac{P_{\text{signal}}}{P_{\text{noise}}}\right) \label{eq:SNR}, 
\end{equation}
where $SNR_{dB}$ represents the target signal-to-noise ratio in decibels, $P_{signal}$ is the power of the ground truth signal, and $P_{noise}$ is the power of noise.
Subsequently, 2,048 Gaussian noise samples, matching the signal length, with mean zero and unit variance were generated and multiplied by the square root of the noise power. This resulted in noisy signals with SNR levels of -20 dB, -15 dB, -10 dB, and -5 dB. An example of ground truth compared to signals corrupted at various noise levels is provided in Fig. \ref{fig:examplesignal}. 



Furthermore, to train the classifier to discriminate between a noisy signal and pure noise, Gaussian noise was also incorporated into the dataset. This increased the dataset size from 63,000 to 126,000 samples, with 50\% corresponding to noisy signals and the remaining 50\% corresponding to pure noise. The data were labelled as 0 for pure noise and 1 for samples with an underlying signal.

\subsection{Data processing}
This stage involved two distinct paths: A) for signal detection and B) for signal extraction. Both paths included two processes: data splitting and Z-score normalisation. In addition, the second path involved performing the single-sided Fast Fourier Transform (FFT) and taking the absolute values of the coefficients before modelling. Both approaches are described in the following subsections, starting with data splitting.

\subsubsection{Data splitting}
Developing robust deep learning models requires partitioning the data into training, validation, and testing sets to mitigate overfitting. Overfitting arises when a model memorises the training data rather than learning meaningful patterns, resulting in excellent performance during training but poor generalisation to unseen data. This discrepancy is often assessed using metrics such as Mean Squared Error (MSE) or Mean Absolute Error (MAE).

\begin{figure}
    \centering
    \includegraphics[width=0.85\linewidth]{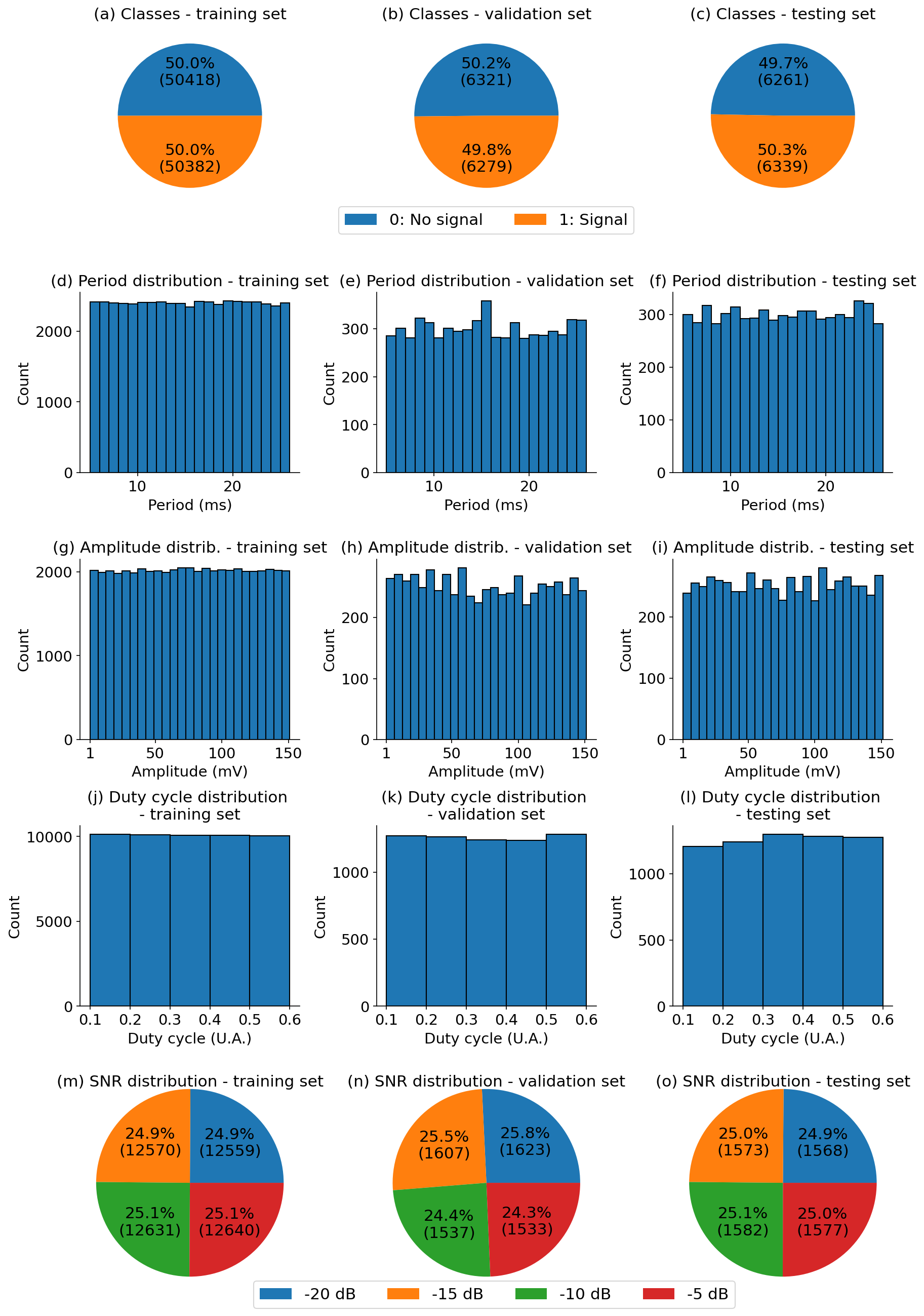}
    \caption[Distributions of the data characteristics across datasets.]{Distributions of the data characteristics across datasets. The class balance of training, validation and testing sets are illustrated in (a), (b), and (c). Period distributions for the training, validation, and testing sets are shown in (d), (e), and (f), while amplitude distributions are depicted in (g), (h), and (i) for the respective sets. Duty cycle distributions are illustrated in (j), (k), and (l), and SNR distributions are displayed in (m), (n), and (o) for the mentioned datasets.}
    \label{fig:stats_sets}
\end{figure}

In this study, the data were split into proportions of 80\%-10\%-10\% for training, validation, and testing, respectively, aligning with previous research \cite{lin2023ecg}. These proportions were applied to both classification and regression tasks. Additionally, since unique signals were generated, there was no data repetition either between or within the sets. For classification, the training set comprised 100,800 samples of noisy signals and pure noise, representing 80\% of the total dataset of 126,000 samples. Both the validation and testing sets consisted of 12,600 signals each, accounting for 10\% of the total. The class balance for `signal' (one) and `no signal' (zero) was verified for each dataset, as shown in Fig. \ref{fig:stats_sets} (a - c).

For the regression task, the data distribution quality was assessed by examining the parameter distributions in each set, as depicted in Fig. \ref{fig:stats_sets} (d - l). In this case, only the signals were used, resulting in a training set of 50,400 signals, representing 80\% of the total 63,000 signals. Both the validation and testing sets consisted of 6,300 signals each, constituting 10\% of the total. As shown in Fig. \ref{fig:stats_sets}, the distributions across the different parameters of period, amplitude and duty cycle are relatively uniform. Moreover, approximately 25\% of the data is allocated to the SNR levels of -20 dB, -15 dB, -10 dB, and -5 dB across the three datasets. This guarantees data quality for the modelling stage. 

\subsubsection{Z-score normalisation}
\label{Zscore}
After data splitting, the Z-score normalisation was applied considering the wide range of signal amplitude, from 1 to 150 mV. First, the mean ($\mu_x$) and standard deviation ($\sigma_x$) of each noisy signal ($X$) were calculated and used to normalise the signal using

\begin{equation}
Z = \frac{X - \mu_x}{\sigma_x}, \label{eq:norm}
\end{equation}
where $X$ represents the noisy signal consisting of 2,048 samples. The resulting $Z$ corresponds to the normalised signal. The $\mu_x$ and $\sigma_x$ values per signal were saved for post-processing.
After the Z-score normalisation, the regression path B) proceeded to the data modelling stage. However, the data used for signal detection underwent an additional step according to path A): the Fast Fourier Transform (FFT).

\subsubsection{Single-sided FFT}
For the classification modelling, the FFT was applied to the signals to obtain their coefficients in the frequency domain. The FFT approach was performed as the signal would have a distinct frequency spectrum in comparison to noise, thus easier to classify. Eq. \ref{eq:fft}:
\begin{equation}
F[k] = \sum_{n=0}^{N-1} f[n] \cdot e^{-i \frac{2\pi}{N} kn},
\label{eq:fft}
\end{equation}
describes complex FFT where $F[k]$ is the coefficient at frequency $k$, $f[n]$ is the value of the signal at $n$, $N$ is the total number of points in the signal, $i$ is the imaginary unit, $2\pi/N$ is the angular frequency of each component, and $k$ is the frequency index.
For signals with a length of 2,048, 1,024 coefficients were obtained from the single-sided FFT. The absolute values of these Fourier coefficients were then computed and saved. These values were used as the input for the classifier to determine whether or not there was an underlying signal.

\subsection{Modelling}
This stage was applied independently to both classification and regression, beginning with the definition of the neural network architecture. It involved training the model, tuning hyperparameters, and calculating performance metrics. Subsequently, during the testing phase, each model was evaluated using test data to assess its performance on unseen data. This concluded the stage before proceeding to the final data post-processing. The sub-stages are detailed in the following subsections.

\subsubsection{Defining the CNN architecture}
\begin{figure}[htbp]
    \centering
    \includegraphics[width=\linewidth]{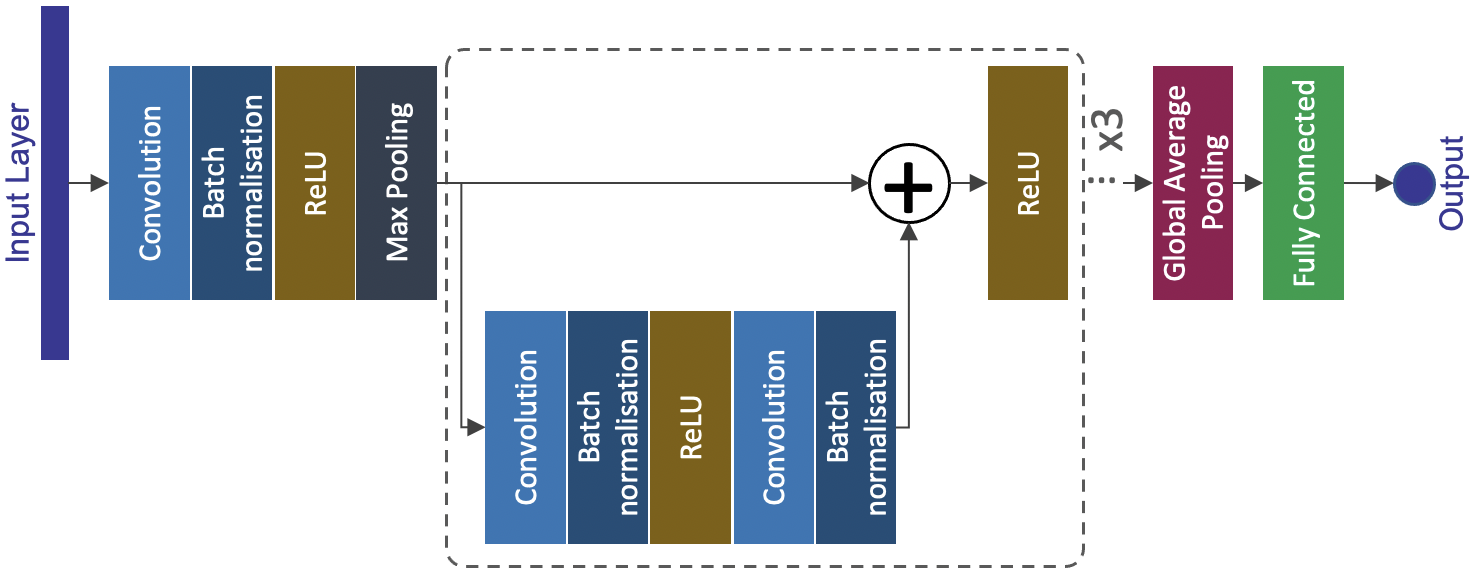}
    \caption[Resnet]{ResNet architecture. This architecture consists of an initial convolutional block, followed by three residual blocks, and concludes with a global average pooling and a fully connected output layer. Each residual block includes a convolutional block with a parallel skip connection that is combined by addition, followed by a ReLU activation layer. The output of the last residual block is followed by a global average pooling, to finish with a fully connected layer of one single unit and a sigmoid activation function that modulates the output between 0 and 1.}
    \label{fig:resnet}
\end{figure}

For signal detection, the classifier was trained on the amplitude of Fourier coefficients. The architecture used for signal detection was a residual network (ResNet) CNN, as illustrated in Fig. \ref{fig:resnet}. This kind of architecture has been widely used to classify various signals, such as arrhythmias from ECG signals \cite{brito2019electrocardiogram, khan2023ecg}, and digital modulation signals in communications \cite{yin2018short, zou2022research}. 

The proposed ResNet, shown in Fig. \ref{fig:resnet}, begins with an initial block consisting of a convolutional layer with 64 filters, a kernel size of 7, and a stride of 2. It is followed by a batch normalisation layer, a ReLU activation, and a max-pooling layer with a pool size of 3 and strides of 2. This initial convolutional block is followed by three residual blocks; each residual block consists of a convolutional layer with 64 filters, kernel size of 3, a batch normalisation, a ReLU activation, and then another convolutional layer followed by batch normalisation. The output of the batch normalisation layer is added to the output from the previous convolutional block carried via skip connection. Afterwards, the residual block ends with a ReLU activation layer. This residual block is repeated three times in sequence. Successive to the ReLU activation from the last residual block, a global average pooling is applied, followed by a fully connected layer with a single unit and a sigmoid activation function that scales the output between 0 and 1. An output value greater than 0.5 indicates the presence of an underlying signal, while a value less than 0.5 suggests that there is no underlying signal, only pure noise.

\begin{figure}[h!]
    \centering
    \includegraphics[width=0.9\linewidth]{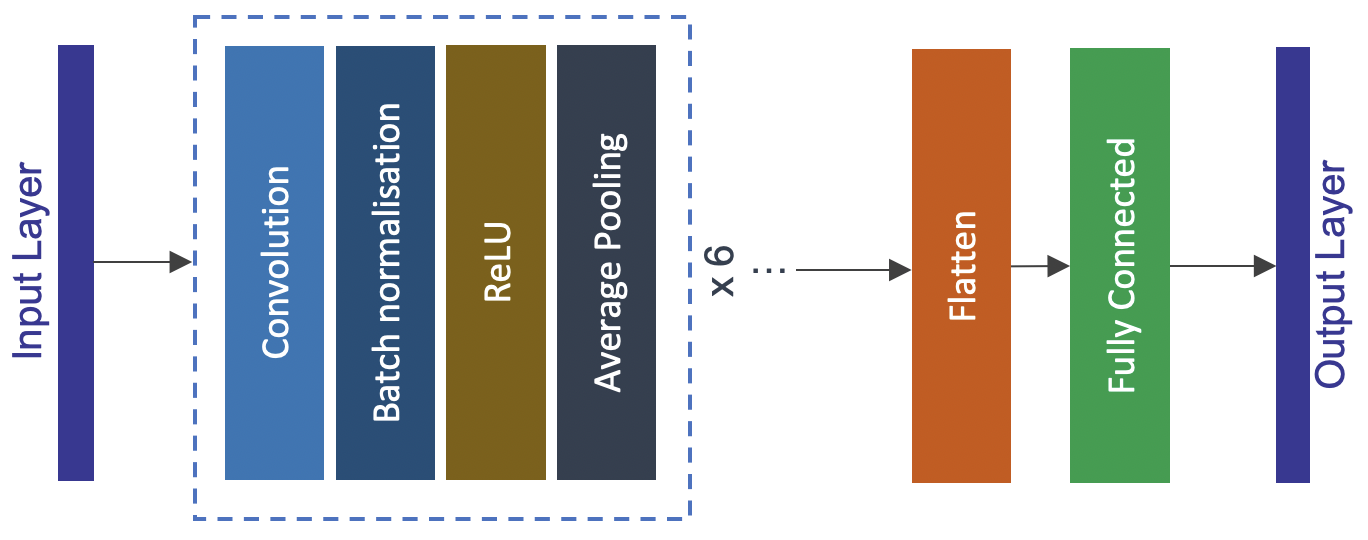}
    \caption[Consecutive convolutional blocks arquitecture.]{Consecutive convolutional blocks (CB-CNN) arquitecture, adapted from \cite{DLECG}. It consists of six blocks, each comprising a convolutional layer, batch normalization, ReLU activation, and average pooling layers. The final steps include a flattening layer, a fully connected layer, and a reshaping layer to restore the input shape.}
    \label{fig:arch1}
\end{figure}

For signal extraction, four CNN architectures were defined and compared. The first CNN architecture is based on previous research by Arsene \textit{et al.} \cite{DLECG} and is depicted in Fig. \ref{fig:arch1}, which was used to process ECG signals. This architecture consisted of six consecutive convolutional blocks; therefore, it is abbreviated here as CB-CNN. The blocks are succeeded by a flatten layer and culminating in a fully connected layer with the same number of units as there are samples (2,048). At the input layer, the data takes the form of 2048x1x1 and the kernels and strides size were adapted according to the data. Each convolutional block comprises a convolutional layer with 36 filters, a kernel size of 19x1, and a stride of 1x1, with zero padding. This is followed by a batch normalisation layer, a ReLU activation layer, and an average pooling layer with 2x1 size and stride of 4x1, without padding. Following the last block, there is a flatten layer and a fully connected layer with 2,048 units, which match the shape of the input signal.

\begin{figure}[h!]
    \centering
    \includegraphics[width=\linewidth]{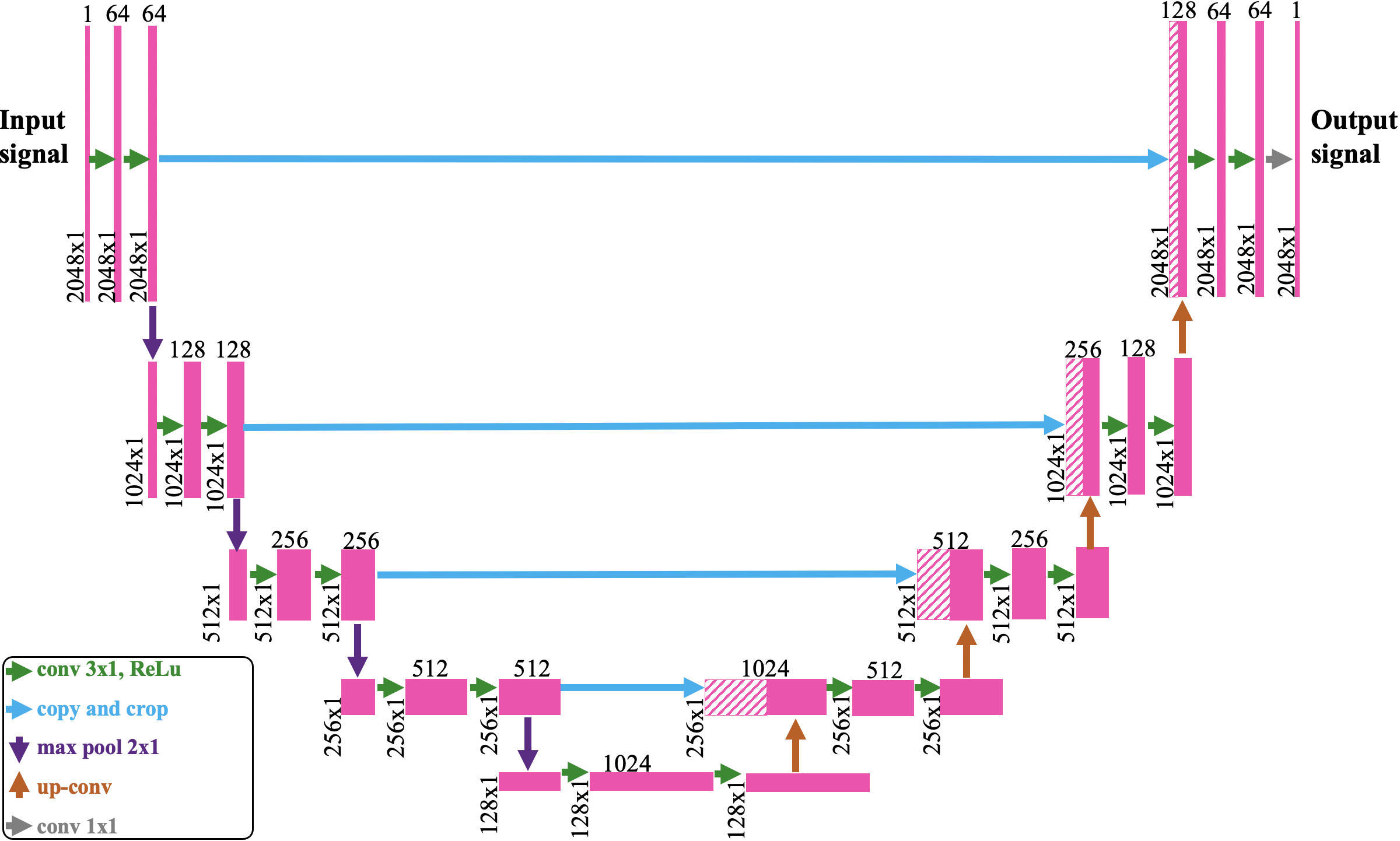}
    \caption[U-Net architecture.]{U-Net architecture, adapted from \cite{ronneberger2022convolutional}. This architecture has four encoding blocks, followed by the bottleneck, and the corresponding four decoding blocks, where information from the encoding and decoding is concatenated for the subsequent signal reconstruction. A final convolution with one filter was performed at the end to restore the shape of the input signal.}
    \label{fig:unet}
\end{figure}

The second architecture is a U-Net structure, which has been employed in previous studies to denoise ECG signals \cite{rasti2022deep} and perform biomedical image segmentation \cite{ronneberger2022convolutional}. This architecture, summarised in Fig. \ref{fig:unet}, comprises a structured arrangement of encoder and decoder blocks, specifically designed for signal processing. The diagram illustrates the U-Net based architecture. It features four encoder blocks with 64, 128, 256, and 512 filters, followed by a bottleneck with two convolutional layers each having 1024 filters. Then, four decoder blocks are arranged with 512, 256, 128, and 64 filters, respectively. A final convolution operation is applied to reduce the dimensionality to 2048, matching the input signal shape. The different operations are represented through green arrows indicating convolution operations with a kernel size of 3x1, zero padding, and ReLU activation. The purple arrows represent max-pooling performed at the end of each block with a pool size of 2x1 and stride of 2x1. The orange arrows depict the upsampling by transposed convolutions with a kernel size of 2x1, strides of 2x1, and zero padding. The sky-blue arrows indicate the copying and cropping of information from feature maps obtained in corresponding encoding blocks. The feature maps concatenation involves pairing half of the upsampled feature maps with their corresponding encoded feature maps. The final convolutional layer uses a single 1x1 kernel filter for signal reconstruction.

\begin{figure}[h!]
    \centering
    \includegraphics[width=0.9\linewidth]{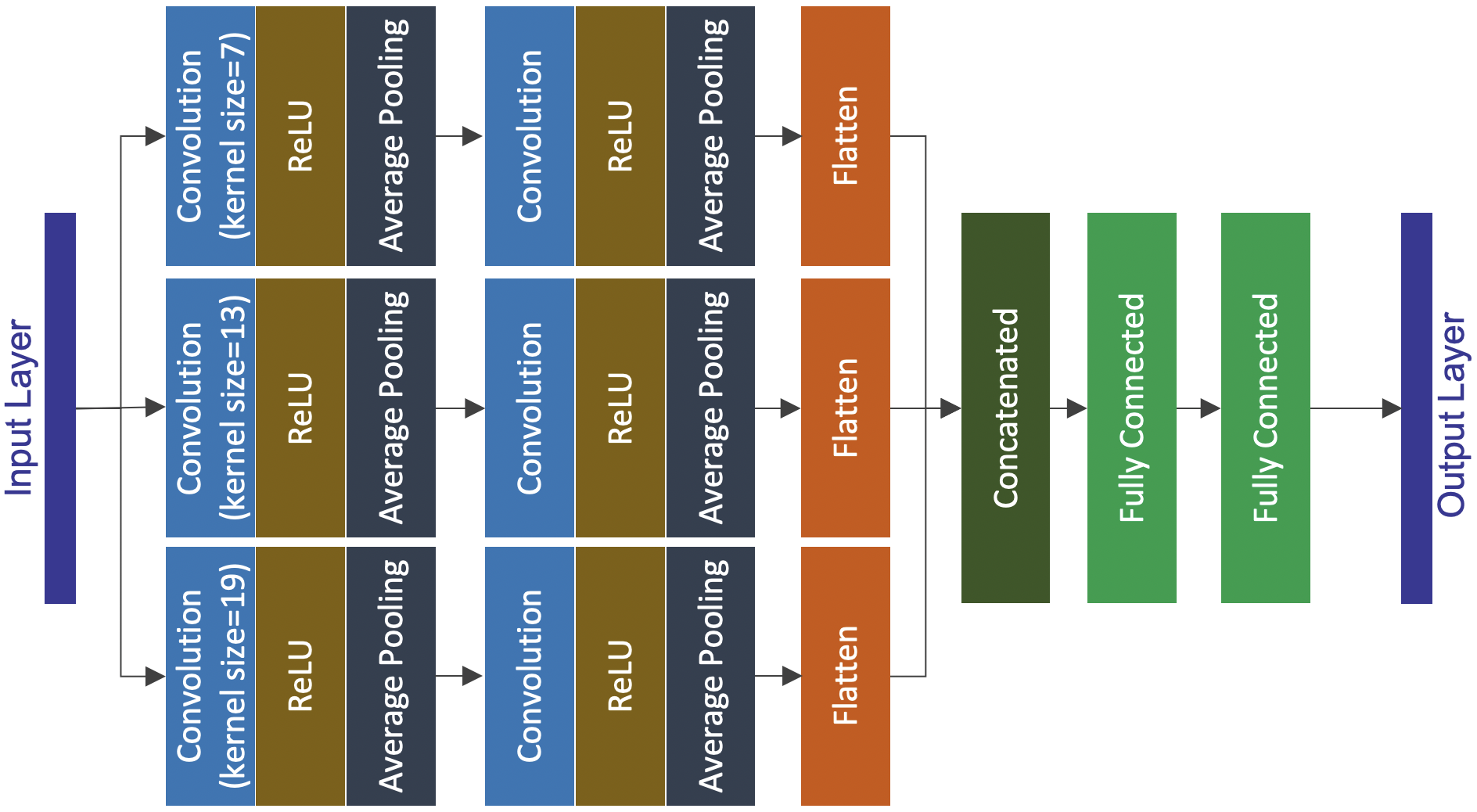}
    \caption[Branched convolutional architecture.]{Branched blocks convolutional (BB-CNN) architecture. It comprises three parallel branches, each consisting of two convolutional blocks followed by a flatten layer. The outputs from these three flatten layers are concatenated, and then, three fully connected layers apply weights to them before reaching the final layer, which matches the number of neurons in the input signal.}
    \label{fig:customcnn}
\end{figure}

The third architecture is inspired by the principles of the Wavelet transform having parallel CNN routes with variable kernel sizes. The concept is demonstrated using an architecture consisting of three parallel branched blocks; therefore, it is abbreviated as BB-CNN. Each branch receives the input signal, as shown in Fig. \ref{fig:customcnn}, and is composed of two convolutional blocks, followed by a flatten layer. The output of the three flatten layers feeds to a concatenation layer which is followed by three consecutive fully connected layers. The first consecutive fully connected layer has 128 units and ReLU activation. The second layer has 256 units and ReLU activation, and the third consists of 2,048 units and linear activation. This results in a final output signal of length 2,048. 

The branches have varying kernel sizes in their convolutional blocks. In the first branch, from the bottom in Fig \ref{fig:customcnn}, the convolutions within the blocks have 32 filters with a kernel size of 19, while in the middle branch, the kernel size is 13, and 7 in the last branch. All of these convolutions have a stride of 1 and are followed by a ReLu activation function. The average pooling operations within each block have a pool size of 2 and stride of 1. Additionally, all operations within blocks include zero padding, resulting in three flatten layers with similar shapes. The various kernel sizes applied are conceptually similar to the scales in the wavelet operation used to convolve the signal of interest. However, instead of fixing a specific wavelet function, it empowers the network to define the filter weights. 

\begin{figure}[ht]
    \centering
    \includegraphics[width=\linewidth]{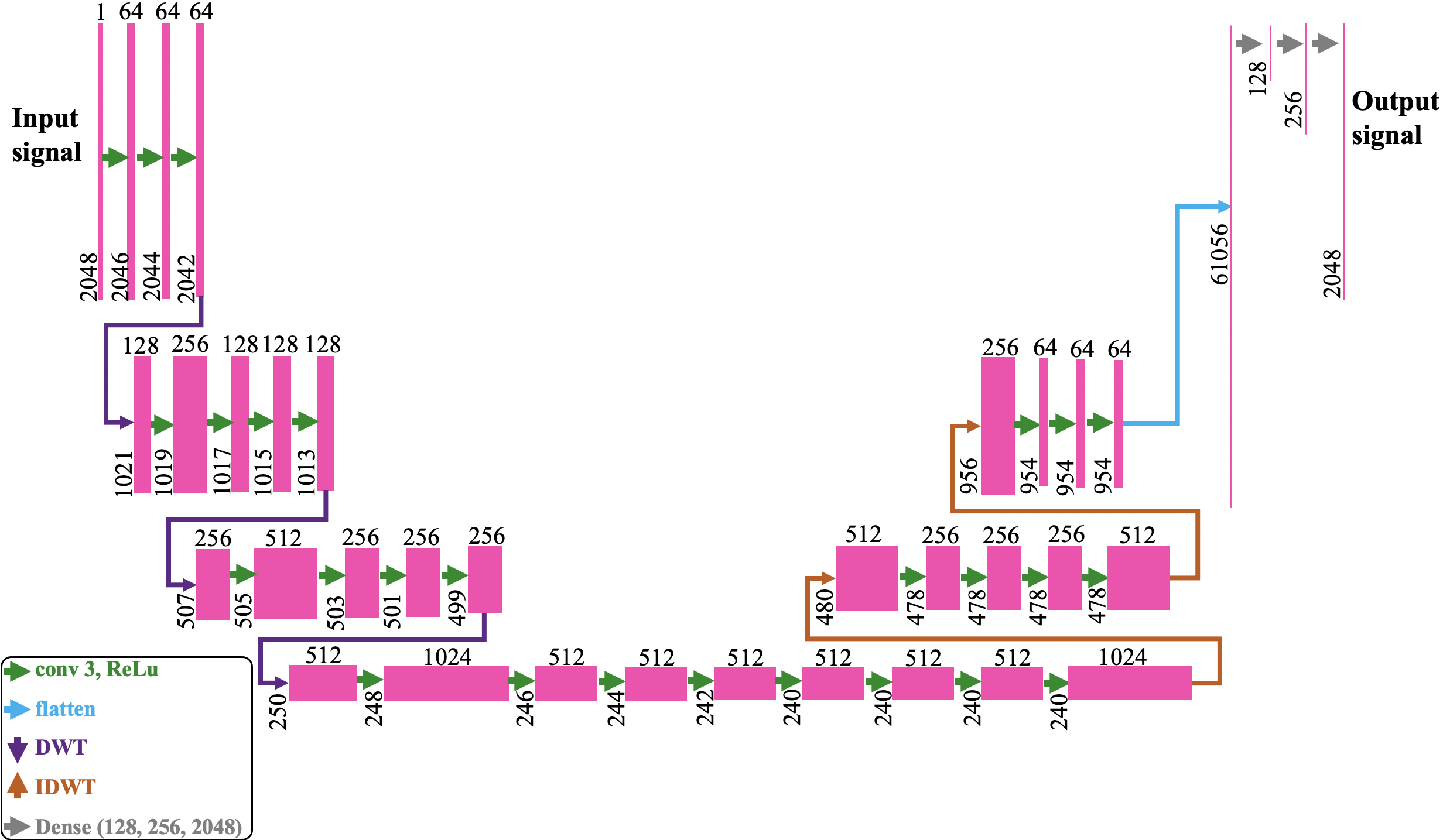}
    \caption[Modified Multilevel Wavelet Convolutional Neural Network architecture.]{Multilevel Wavelet Convolutional Neural Network architecture, adapted from \cite{liu2019multi}. It has encoding and decoding blocks, where green arrows indicate the DWT, while red arrows represent its inverse, replacing the max-pooling and upsampling operations, respectively.}
    \label{fig:mwcc}
\end{figure}

The fourth architecture is a modified Multilevel Wavelet Convolutional Neural Network (MWCNN) model \cite{liu2019multi}. While sharing similarities with U-Net, the MWCNN replaces traditional max-pooling and upsampling operations with the Discrete Wavelet Transform (DWT) and Inverse Discrete Wavelet Transform (IDWT) operations, respectively. Its architecture is illustrated in Fig. \ref{fig:mwcc}.   

The proposed architecture utilises the \verb|WaveTF| library \cite{wavetf} to integrate wavelet-transformed layers into the TensorFlow CNN framework, enabling efficient GPU processing. The architecture begins with three encoding blocks, the first has three 64-filter convolutional layers. The subsequent blocks follow a pattern of a 2n-filter convolutional layer that is succeeded by three n-filter convolutional layers, where n={128, 256}. A DWT operation was applied at the end of each encoding block. Subsequently, a bottleneck of eight convolutional layers is built, with the first layer having 1024, the following six having 512 filters, and the last layer of the bottleneck having 1024 filters. This is followed by two decoding blocks that start with the IDWT. The first decoding block has three 256-filter convolutional layers, followed by a 512-filters convolutional layer before applying the IDWT.  The second block has three 64-filter convolutional layers with no skip connections. A Haar mother wavelet was utilised for performing the DWT and its corresponding IDWT, in the encoding and decoding blocks. All convolutional layers have a kernel size of 3 and a stride of 1, with ReLu activation functions, and are zero-padded in the decoding blocks. Instead of performing another decoding block, a flatten layer is placed, followed by two dense layers of 128 and 256 neurons, each using ReLu activation. The final dense layer, with linear activation, has 2,048 units matching the input data length.

\subsubsection{Training and hyperparameter tuning}
The hyperparameters were initially set and adjusted iteratively based on model performance. A summary of relevant hyperparameters and their values, drawn from related works and adjustments, is presented in Table \ref{table:hyperparameters}.

\begin{table}[ht]
\centering
\footnotesize 
\setlength{\abovecaptionskip}{-0.0pt}
\caption{List of hyperparameters and their values per  architecture.\\}
\setlength\tabcolsep{1.5pt}
\begin{tabular}{|l|c|c|c|c|} 
\hline
 \multicolumn{5}{|c|}{Classification for signal detection} \\ \hline
 Hyperparameter & \multicolumn{4}{c|}{ResNet architecture}\\ \hline
 Learning rate & \multicolumn{4}{c|}{0.0001} \\ \hline
 Optimizer & \multicolumn{4}{c|}{SGD} \\ \hline
 Loss function & \multicolumn{4}{c|}{Binary cross-entropy} \\ \hline
 Batch size & \multicolumn{4}{c|}{8} \\ \hline
 Epochs & \multicolumn{4}{c|}{400} \\ \hline
 Early stopping & \multicolumn{4}{c|}{Val loss, patience=200} \\ \hline
 Metrics & \multicolumn{4}{c|}{Accuracy} \\ \hline
 \multicolumn{5}{|c|}{Regression for signal extraction} \\ \hline
 Hyperparameter & \makecell[tc]{\scriptsize Architecture 1\\ \scriptsize{CB-CNN}} & 
 \makecell[tc]{\scriptsize Architecture 2\\ \scriptsize{U-Net}} & 
 \makecell[tc]{\scriptsize Architecture 3\\ \scriptsize{BB-CNN}} & 
 \makecell[tc]{\scriptsize Architecture 4\\ \scriptsize{MWCNN}} \\
\hline
 Learning rate & \multicolumn{4}{c|}{0.0001} \\ \hline
 Optimizer & \multicolumn{4}{c|}{Adam} \\ \hline
 Loss function & \multicolumn{4}{c|}{MSE} \\ \hline
 Batch size & \multicolumn{1}{c|}{8} & \multicolumn{1}{c|}{16} 
 & \multicolumn{1}{c|}{8} & \multicolumn{1}{c|}{16}\\ \hline
 Epochs & \multicolumn{4}{c|}{40}\\ \hline
 Early stopping & \multicolumn{4}{c|}{Val loss, patience=5} \\ \hline
 Metrics & \multicolumn{4}{c|}{MAE, RMSE}\\ \hline
 \makecell[tl]{\scriptsize Wavelet function layer} & \multicolumn{3}{c|}{N/A} & Haar \\
\hline
\end{tabular}
\label{table:hyperparameters}
\end{table}

In the implementation of the described architectures, the influence of the following was investigated: setting the MAE as the loss function, adding batch normalisation layers to the convolution blocks, defining a custom loss function based on the SNR, another custom loss based on cosine similarity, combining it with the MSE, changing the activation functions to Leaky ReLU, and adjusting the padding in different layers. The adopted hyperparameters were selected based on model performance, as explained in the following subsection.

\subsubsection{Metrics}
The model performance was assessed using various metrics described below. For training the classifier, the binary cross-entropy (BCE) \cite{bishop2006} was used as the loss function, represented by
\begin{equation}
BCE = - \frac{1}{N} \sum_{i=1}^{N}[y_i \cdot \log(\hat{y_i}) + (1 - y_i) \cdot \log(1 - \hat{y_i})],
\label{eq:BCE} 
\end{equation}
where $y_i$ is the true label of the i-th instance (0 for no signal and 1 for signal) and $\hat{y_i}$ is the predicted probability of the the i-th instance being in class 1. Therefore, for a 0-class instance, the BCE is given by -log(1-probability of being class 1), while for a 1-class instance is given by -log(probability of being class 1). For a class 0, ideally the probability of being class 1 would be zero, thus, the BCE would be minimised. Conversely, if that probability is close to 1, failing the class prediction, the BCE increases.
Since the probabilities of being in class 0 and class 1 are complementary, the BCE loss increases when instances that are actually class 0 are predicted with a high probability of being class 1, and vice versa. 

Additionally, accuracy was calculated during the training, validation and testing, as the datasets are balanced. The accuracy measures the number of correct predictions over the total, which can be calculated using

\begin{equation}
Accuracy= \frac{TP+TN}{TP+TN+FP+FN},
\label{eq:Acc} 
\end{equation}
where $TP$ (True Positives) are the instances where both the predicted class and the true class are positive (class 1). $TN$ (True Negatives) are the instances where both the predicted class and the true class are negative (class 0). $FP$ (False Positives) occur when the predicted class is positive (class 1), but the true class is negative (class 0). $FN$ (False Negatives) occur when the predicted class is negative (class 0), but the true class is positive (class 1). In summary, $TP$ and $TN$ indicate correct predictions, while $FP$ and $FN$ indicate incorrect predictions. Furthermore, the confusion matrix including the $TP$, $TN$, $FP$ and $FN$ was calculated on the testing set to evaluate model performance. Additionally, the precision and recall were computed, as defined in Eq. \ref{eq:Precision} and Eq. \ref{eq:Recall}:

\begin{equation}
Precision= \frac{TP}{TP+FP},
\label{eq:Precision} 
\end{equation}

\begin{equation}
Recall= \frac{TP}{TP+FN}
\label{eq:Recall} 
\end{equation}

Precision measures the fraction of the correctly predicted positives over all positive predictions regardless of their correctness. Furthermore, recall measures the fraction of the correctly predicted positives over all actual positive instances.

Regarding, the regression for signal extraction, the weights of the model were adjusted based on training and validation loss function values. In this case, the loss function was the MSE, defined in Eq. \ref{eq:MSE}. The RMSE, defined by Eq. \ref{eq:RMSE}, and the MAE, defined by Eq. \ref{eq:MAE}, were also calculated to monitor performance and assess generalisation. 

\begin{equation}
MSE = \frac{1}{n} \sum_{i=1}^{n} (y_i - \hat{y_i})^2 \label{eq:MSE} 
\end{equation}

\begin{equation}
RMSE = \sqrt{\frac{1}{n} \sum_{i=1}^{n} (y_i - \hat{y}_i)^2} \label{eq:RMSE} 
\end{equation}

\begin{equation}
MAE = \frac{1}{n} \sum_{i=1}^{n} |y_i - \hat{y_i}|
 \label{eq:MAE} 
\end{equation}

The model training persisted until the MSE value on the validation set exhibited a consistent plateau lasting for five consecutive epochs. To avoid overfitting, this training should be terminated if there is a consistent plateau or increase in the validation loss lasting for five consecutive epochs.

Afterwards training, the test data were fed into the model to estimate the underlying signals. The previous metrics were calculated and the output SNR was computed using Eq. \ref{eq:SNR} relative to the input SNR to calculate the improvement, as expressed in Eq. \ref{eq:SNRimp}.

\begin{equation}
SNR_{imp\ dB} = SNR_{out\ dB} - SNR_{in\ dB}
 \label{eq:SNRimp} 
\end{equation}

This metric required both the input and output of the model to be on the same scale. Hence, a final post-processing transformation of the inverse Z-score normalisation was applied, as explained in the following section.

\subsection{Data post-processing}
Different post-processing transformations were applied for classification and regression. For classification, the signal and no signal classes were generated based on the output from the single neuron modulated by the sigmoid function. Values below 0.5 were classified as 0 (no signal), whereas values from 0.5 to 1 were classified as 1 (signal). Subsequently, the confusion matrix was calculated.

Regarding regression for denoising, the models were trained using Z-score normalised data. Therefore, the outputs of the model were re-scaled by using the mean ($\mu_x$) and standard deviation ($\sigma_x$) from the data processing stage. The final signal was obtained through the Eq. \ref{eq:norm_inverse}.
\begin{equation}
Y_{final} = {Y_{pred} \ \sigma_x} + \mu_x \label{eq:norm_inverse},
\end{equation} 
here, $Y_{pred}$ represents the signal estimated by the model, and $Y_{final}$ is the signal after the post-processing.
The final formatting involved ensuring that the data matched the time domain signal, resulting in 2,048 samples per signal of a total of 0.1024 seconds duration.

\section{Results}
This section presents the key findings of our investigation of deep CNN-based classification and regression, with a particular focus on their effectiveness in the detection and extraction of target signals under noisy measurement conditions. We present and discuss the ability to correctly predict the presence or absence of the target signal with different SNR levels between - 20 and -5 dB. This is followed by a comparison of the effectiveness of various CNN architectures in signal retrieval against ground truth. 

\subsection{Signal detection}
\begin{figure}[ht] 
    \centering
    \includegraphics[width=0.5\linewidth]{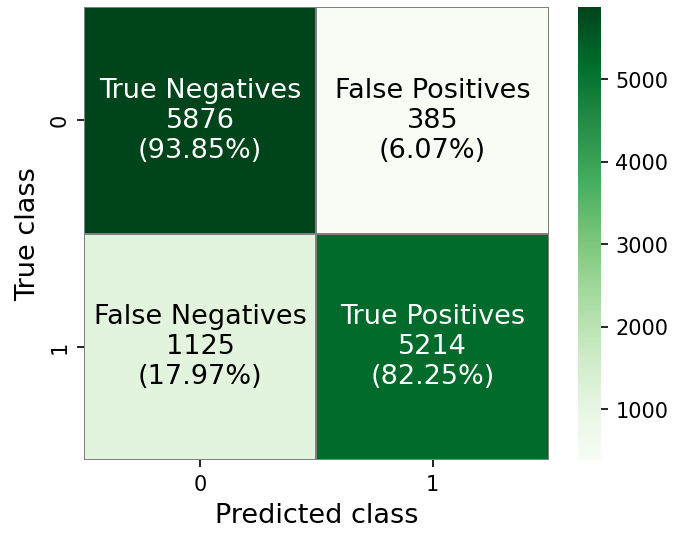}
    \caption[Confusion matrix]{Confusion matrix showing the results of the ResNet binary classifier. The diagonal elements represent the correct predictions, with the top right indicating the number of instances predicted as class 1 but actually belonging to class 0 (false positives), while the bottom left represents the instances predicted as class 0 but actually belonging to class 1 (false negatives).}
    \label{fig:conf_matrix}
\end{figure}

\begin{figure}[t!]
    \centering
    \includegraphics[width=0.9\linewidth]{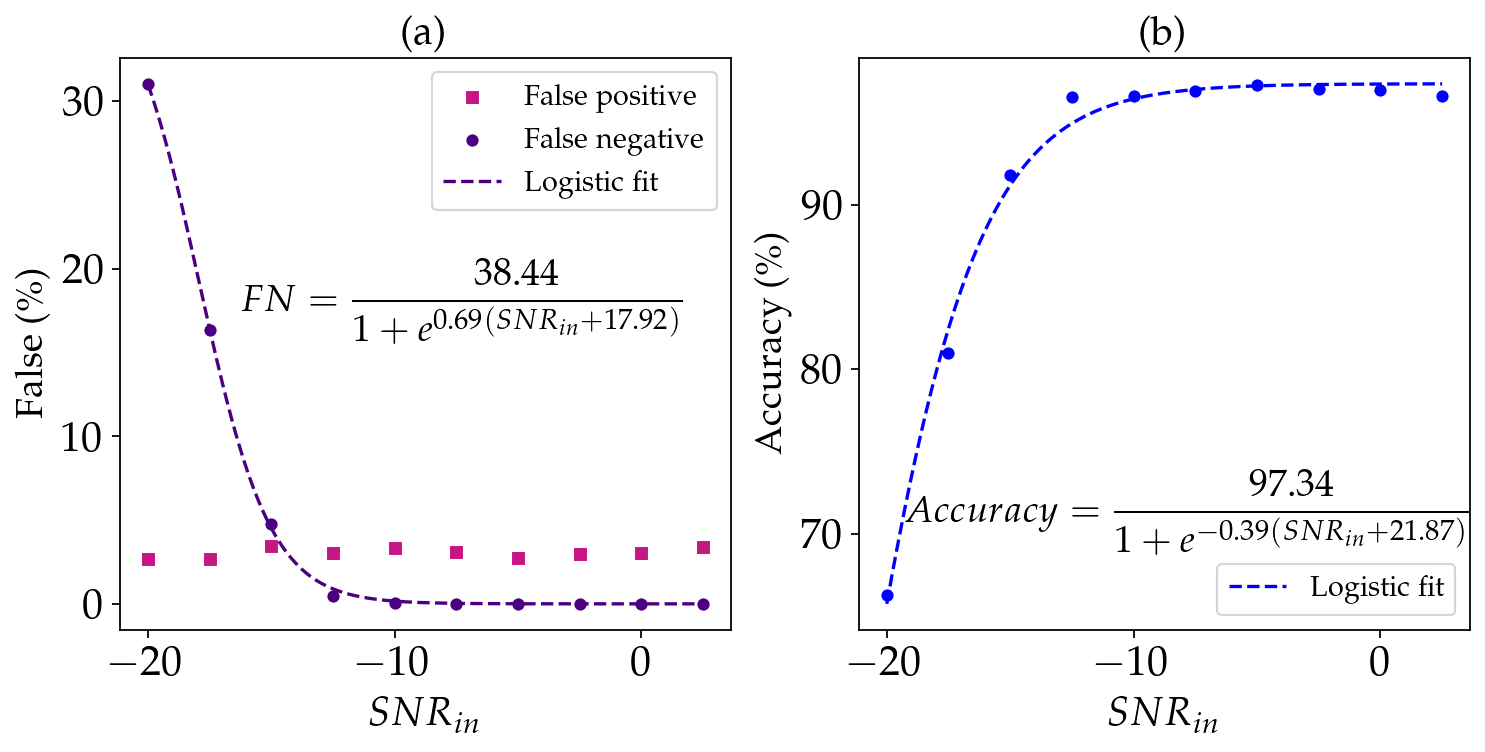}
    \caption[Metrics on classifier]{Metrics of the ResNet binary classier at different input SNR levels. Panel (a) displays the false negative and false positive percentages, while panel (b) shows the accuracy percentage  obtained from the model.}
    \label{fig:accuracy}
\end{figure}

In this section, we present the results of testing the ResNet model for signal detection. Here we achieve an accuracy of 88\% for input SNR ranging from -5 to -20 dB, as calculated using Eq. \ref{eq:Acc}. In other words, a total of 11,090 out of 12,600 samples were correctly classified. The confusion matrix is shown in Fig \ref{fig:conf_matrix}, presenting the correct prediction, false negatives and false positives. The model performed well in predicting class 0, with a true negative rate of 93.85\%. However, while the model identified 82.25\% of the positive cases, it has a false negative rate of 17.97\%, which reflects the proportion of missed positive instances. Furthermore, when investigating these results according to the corresponding input SNR, we observe that less corrupted signals are more correctly classified when compared to more corrupted signals, as shown in Fig. \ref{fig:accuracy} a. For SNR greater than -10 dB, 0\% of false negative was achieved, while it exhibits logistic growth for input SNR of less than -10 dB with 31\% for SNR of -20 dB. On the other hand, the false positives are approximately 3\% for the input SNR levels between -20 and 2.5 dB. Therefore, the accuracy follows a similar trend as presented in Fig. \ref{fig:accuracy} b.

Precision and recall, described in Eq. \ref{eq:Precision} and \ref{eq:Recall}, are widely used metrics for assessing the performance of classifiers. Precision measures the proportion of true positives over all the instances that were classified as positive. In the confusion matrix, it is equivalent to the true positives obtained by performing a vertical sum of instances classified as 1. On the other hand, recall, also known as sensitivity, measures the proportion of true positives that were correctly identified out of all actual positive ground truths. In the confusion matrix, this corresponds to the true positives obtained through the horizontal sum of true class 1 (i.e., false negatives and true positives). Using these metrics, the ResNet classifier achieved a precision of 93\%, representing the percentage of the instances that are correctly predicted as signals (i.e., correspond to the positive class). Furthermore, the model's recall is 82\% which quantifies its ability to capture the  actual positive instances (i.e., signal). Since the false negatives increase as the input SNR decreases, the overall recall of the model is lower than its precision.

 A recently published study has tested LSTM for detecting signals in low SNR conditions \cite{lacy2024machine}. The study used a Gaussian-shaped signal with specific amplitude and duration as a ground truth. This LSTM classifier achieved a high accuracy of 98\% for SNR of -12 dB and 75.2\% for SNR of -21.6 dB. Our deep CNN model has achieved a comparable performance, as shown in Fig. \ref{fig:accuracy} a, with an accuracy of 96.5\% and 66.3\%, for input SNR of -12.5 dB and -20 dB, respectively. The outcomes of this classification can inform the next step of the CNN-based regression. For instance, given the negligible false negative for SNR greater than -10 dB, as shown in Fig. \ref{fig:accuracy} a, the regression step can be bypassed when no signal is detected. 
 These results are promising when considering the detection of fast dynamic bioelectrical signals that are widely observed in living systems. Here we showcase the ability to classify signals with a rising time of 50 $\mu$s while training and testing the deep learning model using signals with variable amplitudes, duty cycles and periods.

\subsection{Signal extraction}
\begin{figure}[ht!]
    \centering
    \includegraphics[width=0.6\textwidth]{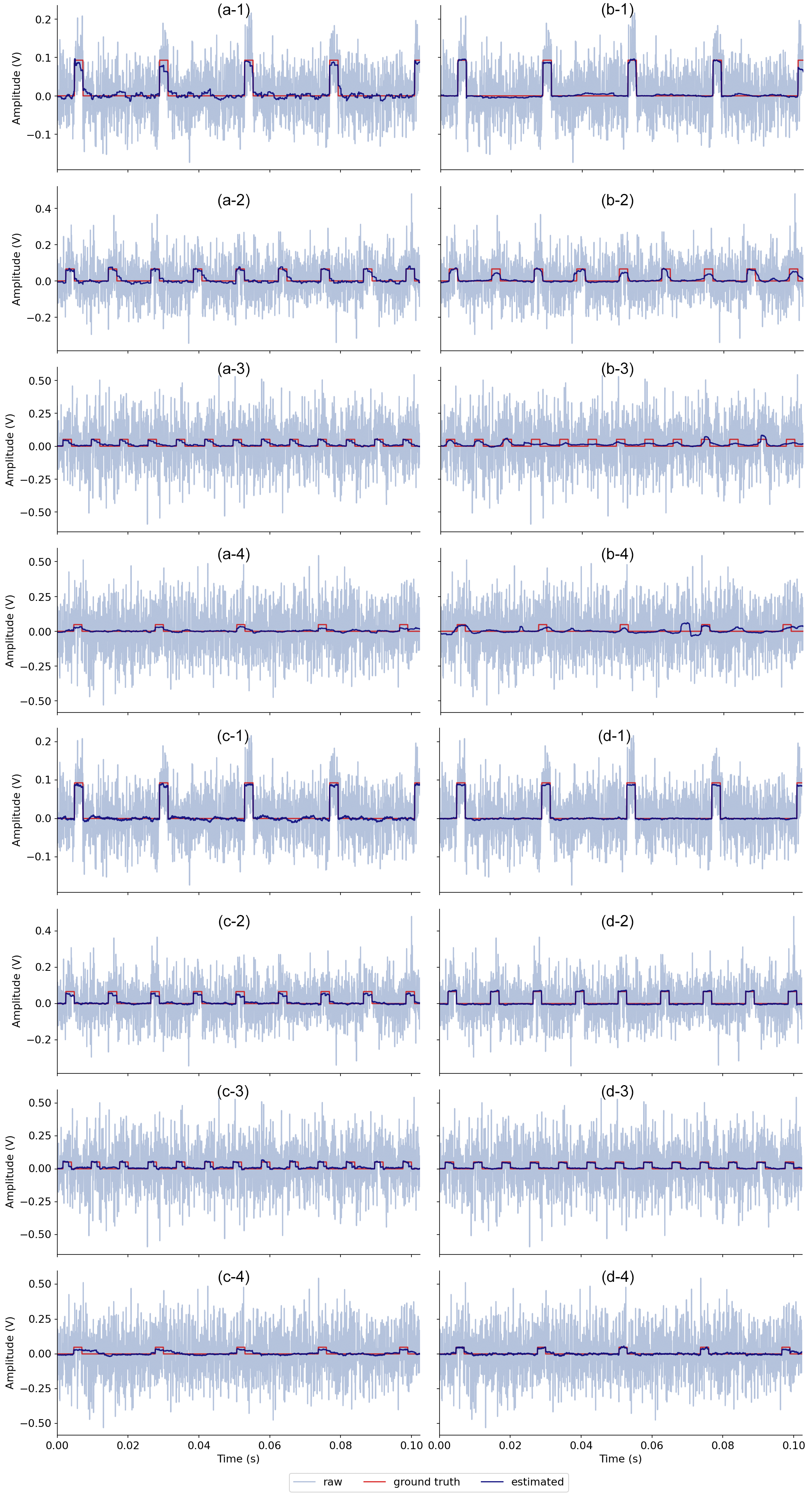}
    \caption[Examples]{Examples of signals with average $SNR_{out}$ from CB-CNN architecture (a), U-Net architecture (b), BB-CNN architecture (c), and MWCNN architecture (d). The signals in the first row correspond to an $SNR_{in}$ of -5 dB, while the signals in the second, third, and fourth rows correspond to $SNR_{in}$ values of -10 dB, -15 dB, and -20 dB, respectively. In each case, the signal in light blue corresponds to the raw signal to be denoised, the ground truth signal is presented in red, and the estimated signal is presented in blue.}
    \label{fig:regular_examples}
\end{figure}

This section presents the performance of the four CNN architectures described in Section \ref{methods}. 
As presented in Table \ref{table:all_metrics_aggregated}, the MWCNN architecture has shown the best performance with an average improvement in SNR of 26 $\pm$ 3.2 dB. The performance varied among the $SNR_{in}$ levels in the range of -5 to -20 dB with a decrease in performance as $SNR_{in}$ decreases. Signals with an $SNR_{in}$ of -5 dB improved to 22.9 $\pm$ 3.1 dB, whereas signals with an $SNR_{in}$ of -10 dB improved to 16.7 $\pm$ 3 dB. When the $SNR_{in}$ decreased to -15 dB the signals reached an average $SNR_{out}$ of 10 $\pm$ 2.9 dB. Further lowering the $SNR_{in}$ to -20 dB resulted in an average $SNR_{out}$ of 4.5 $\pm$ 2.5 dB. Examples of signal regression are displayed in Fig. \ref{fig:regular_examples}. Better estimates were achieved for higher $SNR_{in}$ levels. The performance varies among the four CNN architectures with apparent improvement in shape estimation when testing MWCNN. The performance metrics of the four architectures in denoising signals, and the effect of the signal parameters on the performances are presented in the following subsections.

\subsubsection{Performance metrics}
\begin{figure}[ht!]
    \centering
    \includegraphics[width=0.9\linewidth]{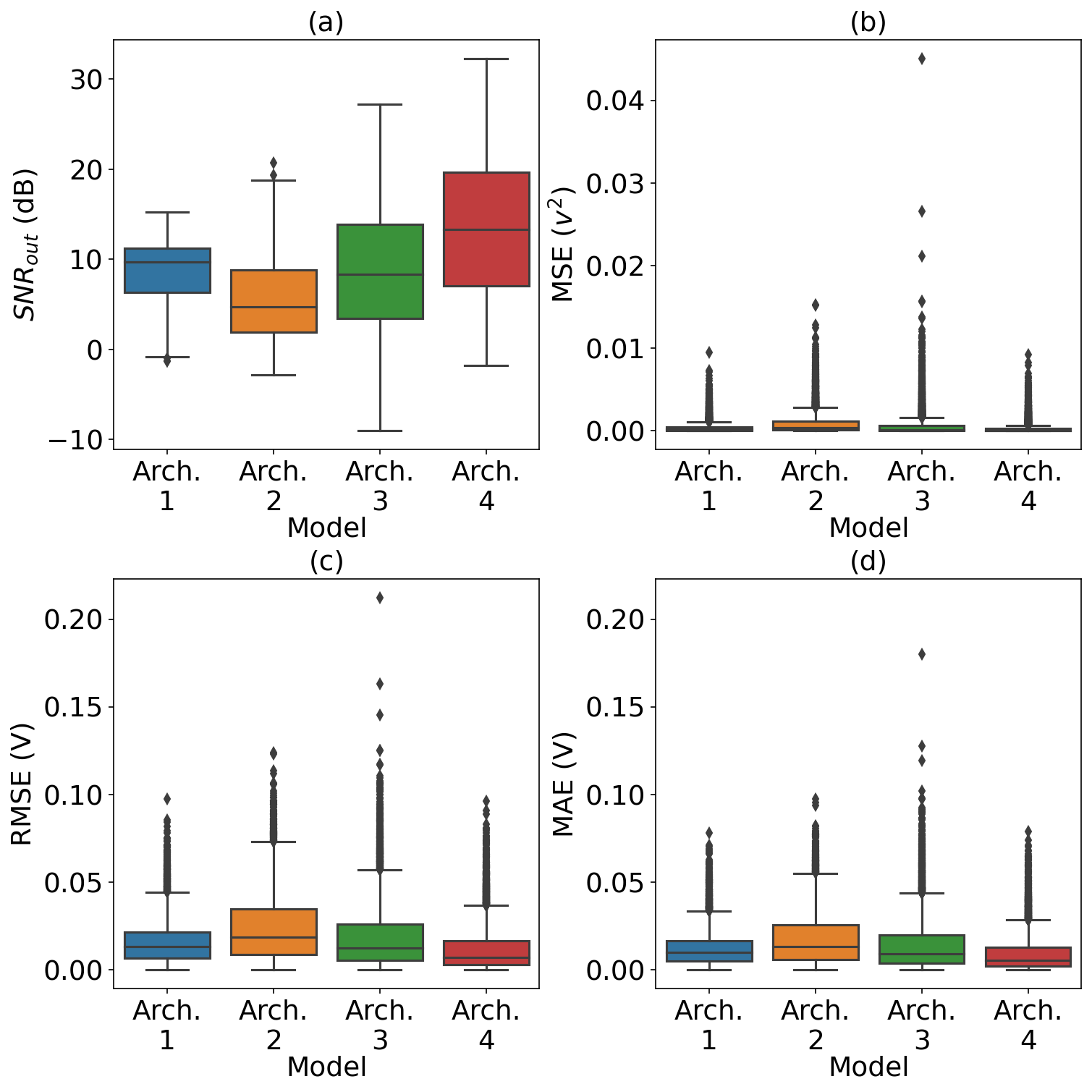}
    \caption[Metrics distributions per architecture.]{Metrics distributions per architecture. The $SNR_{out}$, MSE, RMSE and MAE are shown in (a), (b), (c) and (d). This includes signals of $SNR_{in}$ of -5 dB, -10 dB, -15 dB, and -20 dB. Distributions are illustrated in blue, orange, green and red, for architecture 1 (CB-CNN), 2 (U-Net), 3 (BB-CNN), and 4 (MCWNN) architectures, respectively.}
    \label{fig:all_metrics_general}
\end{figure}

To assess model performance for signal denoising, the MSE, RMSE, MAE, and SNR metrics were calculated. The four architectures were evaluated by calculating these metrics using a test set of 6,300 raw noisy signals. The distributions of the metrics are shown in Fig. \ref{fig:all_metrics_general}.

As illustrated in Fig. \ref{fig:all_metrics_general} (a), the MWCNN architecture (i.e., architecture 4) achieved the highest median $SNR_{out}$ of 13.34 dB. It is followed by architectures 1 (CB-CNN), 3 (BB-CNN), and 2 (U-Net) with medians of 9.70 dB, 8.31 dB, and 4.77 dB, respectively (refer to Table \ref{table:all_metrics_aggregated}).
 Notably, the trend observed in Fig. \ref{fig:all_metrics_general} (a) is consistent with the error metrics, showed in Fig. \ref{fig:all_metrics_general} (b), (c), and (d); the MWCNN architecture exhibited the lowest median MSE, RMSE, and MAE with $4.7$ $10^{-5} (mV)^2$, $6.9\ \mu V$ and $5.5\ \mu V$, respectively. As shown in Table \ref{table:all_metrics_aggregated}, the mean RMSE obtained from the MWCNN was 0.0119 mV. This value shows a modest decrease compared to the reported average RMSE of 0.01355 mV in related works, where ECG signals with an amplitude of 1 mV, affected by electrode motion noise, and with an input SNR of 0 dB were processed \cite{lin2023ecg}. Similarly, the mean RMSE of the CB-CNN architecture was 0.0160 mV, being marginally lower than the reported 0.01869 mV of average RMSE after processing ECG signals with an amplitude of 1 mV, affected by baseline wander noise, and with an input SNR of 0 dB \cite{lin2023ecg}. The architectures U-Net and BB-CNN exhibited comparable RMSE to the literature \cite{lin2023ecg}, with 0.0237 mV and 0.0188 mV, respectively. 
 
 Moreover, the proposed MWCNN architecture achieved an average SNR improvement of 26 dB across all studied input SNR of -20 dB, -15 dB, -10 dB, and -5 dB. This improvement is modestly higher that the SNR improvement reported in previous studies, where ECG signals normalised to one unit of amplitude and with an input SNR of -1 dB, affected by realistic mixed noises (baseline wander, muscle artifacts and electrode motion noises), showed an SNR improvement of approximately 21 dB \cite{rasti2022deep}. The same study reported an RMSE of 0.034 (normalised scale) which is higher than the mean RMSE of the four previously mentioned architectures.
 

\begin{table}[htbp]
  \centering
  \caption{$SNR_{out}$ for different architectures.}
  \renewcommand{\arraystretch}{0.7}
  \setlength{\tabcolsep}{0.7pt}
  \small
  \begin{tabular}{|ll|r|r|r|r|r|}
    \toprule
    Metric & & 
    \makecell[tc]{\scriptsize Architecture 1\\ \scriptsize{CB-CNN}}  & \makecell[tc]{\scriptsize Architecture 2\\ \scriptsize{U-Net}} &  \makecell[tc]{\scriptsize Architecture 3\\ \scriptsize{BB-CNN}} & 
    \makecell[tc]{\scriptsize Architecture 4\\ \scriptsize{MWCNN}}  \\
    \cmidrule(lr){1-6}
    $SNR_{out}$ & \scriptsize{Median} & 9.6981 & 4.7690 & 8.3118 & \textbf{13.3438} \\
    (dB) & \scriptsize{Mean} & 8.7365 &  5.6267 &  8.9062 & \textbf{13.5343} \\
             & \scriptsize{Std} & \textbf{3.4244} &  4.5404 &  6.4317 &  7.4833 \\
             & \scriptsize{CV} &  \textbf{0.3920} & 0.8069 & 0.7222 & 0.5529 \\
    \cmidrule(lr){1-6}
    $SNR_{imp}$ & \scriptsize{Median} & 21.4561 & 18.1825 & 21.2133 & \textbf{25.8595} \\
     (dB)    & \scriptsize{Mean} & 21.2222 & 18.1124 & 21.3919 & \textbf{26.0200} \\
             & \scriptsize{Std} & 3.2892 &  \textbf{2.2965} &  3.5946 &  3.1859 \\
             & \scriptsize{CV} &  0.1550 & 0.1268 & 0.1680 & \textbf{0.1224} \\
    \cmidrule(lr){1-6}
    MSE & \scriptsize{Median} & 0.0002 &  0.0003 &  0.0002 &  \textbf{0.0000} \\
    (mV)$^2$  & \scriptsize{Mean} & 0.0004 &  0.0009 &  0.0007 &  \textbf{0.0003} \\
             & \scriptsize{Std} & 0.0007 & 0.0014 & 0.0016 & \textbf{0.0007}  \\
             & \scriptsize{CV} &  1.6853 & \textbf{1.5258} & 2.2018 & 2.2234 \\
    \cmidrule(lr){1-6}
    RMSE & \scriptsize{Median} &  0.0132 & 0.0187 & 0.0124 & \textbf{0.0069} \\
    (mV)       & \scriptsize{Mean} &  0.0160 & 0.0237 & 0.0188 & \textbf{0.0119} \\
              & \scriptsize{Std} & \textbf{0.0128} & 0.0191 & 0.0190 & 0.0132 \\
              & \scriptsize{CV} &  \textbf{0.8010} & 0.8064 & 1.0138 & 1.1065 \\
    \cmidrule(lr){1-6}
    MAE & \scriptsize{Median} & 0.0100 &  0.0131 &  0.0092 &  \textbf{0.0055} \\
    (mV)     & \scriptsize{Mean} & 0.0124 &  0.0175 &  0.0146 &  \textbf{0.0095} \\
             & \scriptsize{Std} & \textbf{0.0104} &  0.0151 &  0.0154 &  0.0106 \\
             & \scriptsize{CV} &  \textbf{0.8384} & 0.8644 & 1.0543 & 1.1253 \\
    \bottomrule
  \end{tabular}
  \label{table:all_metrics_aggregated}
\end{table}

To assess the dispersion of metrics such as output SNR, both the standard deviation (Std) and the coefficient of variation (CV) were used. The standard deviation provides a measure of absolute variability, while the CV, which is the ratio of the standard deviation to the mean, offers insight into relative variability. As shown in Table \ref{table:all_metrics_aggregated}, architectures 1 and 2, namely CB-CNN and U-Net architectures, demonstrated  a standard deviation of the output SNR below 4.55 dB. Conversely, the BB-CNN and MWCNN architectures, exhibited higher standard deviations of the output SNR, at 6.4 dB and 7.5 dB, respectively. Specifically, the MWCNN architecture demonstrated the highest variance in output SNR, but it also showed the best relative variability in SNR improvement, with a CV of 0.12. Furthermore, the MWCNN exhibited the lowest standard deviation of the MSE at 0.0007 $(mV)^2$. Along with the CB-CNN architecture, these two architectures achieved the lowest standard deviations and CV for both the RMSE and MAE metrics, demonstrating less dispersion in these error metrics.

Compared to the literature, the standard deviation of the output SNR in this study is higher than the reported values, which range from 1.17 to 2.07 dB \cite{lin2023ecg}. The standard deviation of the RMSE in this work varied from 0.0128 to 0.0191 mV, which is comparable to the reported values of 0.0052 to 0.0177 mV \cite{lin2023ecg}. However, it is noteworthy that in this study the models were evaluated on signals with amplitudes ranging from 1 to 150 mV, whereas the ECG signals reported in the literature were normalised to a fixed voltage range of [0, 1] mV \cite{lin2023ecg}.


\subsubsection{Effect of signal parameters}
\begin{figure}[ht!]
    \centering
    \includegraphics[width=0.7\linewidth]{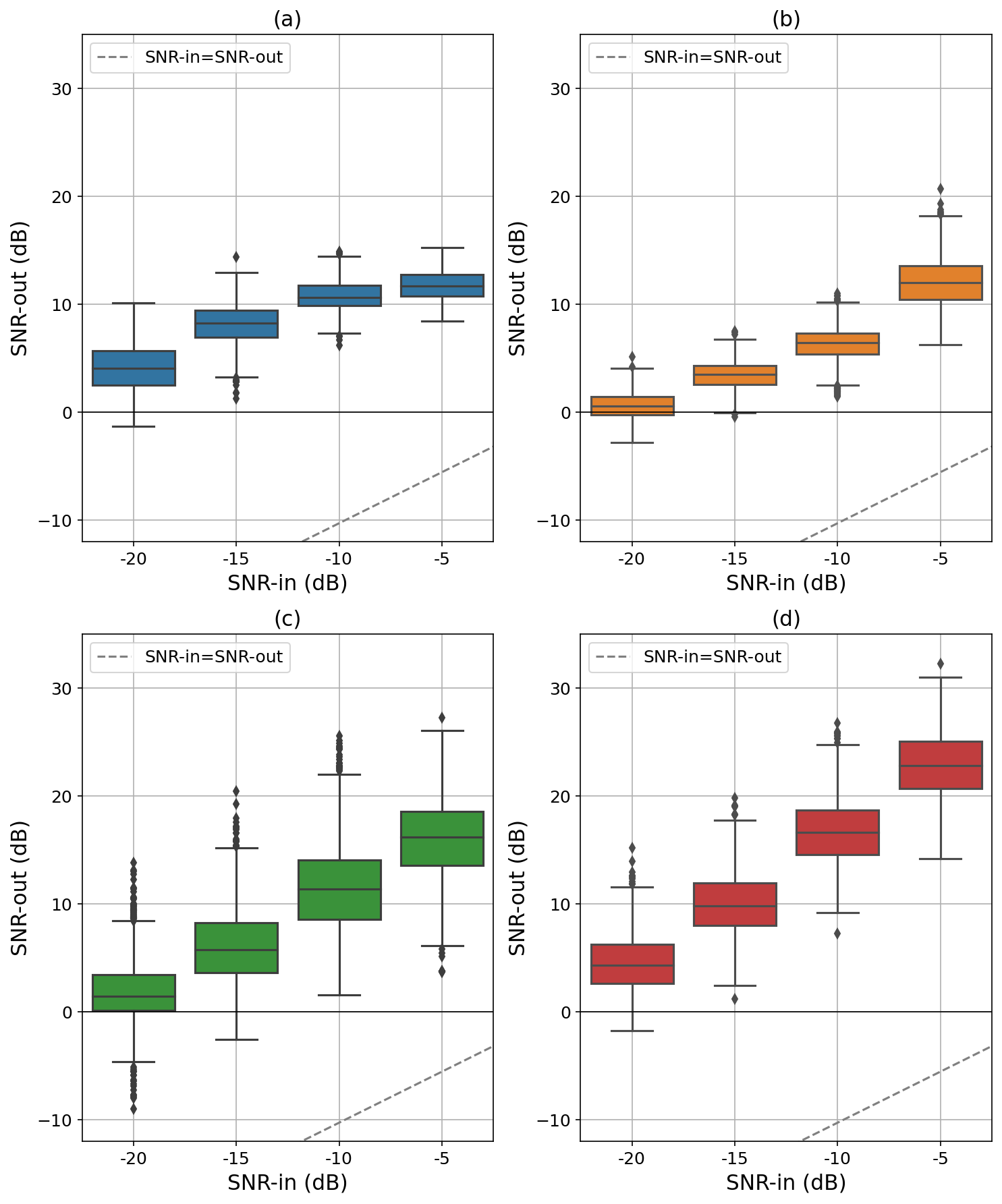}
    \caption[$SNR_{out}$ against $SNR_{in}$ boxplot.]{Output SNR as a function of input SNR levels. The identity line represents the cutoff where $SNR_{in}$ equals $SNR_{out}$, with signals above the line indicating those enhanced by (a) architecture 1 (CB-CNN), achieving a higher $SNR_{out}$. The same information is shown in panels (b), (c), and (d) for architectures 2 (U-Net), 3 (BB-CNN), and 4 (MWCNN), respectively.}
    \label{fig:SNR_out_in_boxplot_all}
\end{figure}

In this subsection, we present and discuss the performance of the different architectures considering the output SNR as a function of the input SNR, and across the ranges of period (5-25 ms), amplitude (1-150 mV) and duty cycle (0.1-0.5).
First, the output SNR is presented against the corresponding input SNR for each architecture as shown in Fig. \ref{fig:SNR_out_in_boxplot_all}. This figure shows that the architectures exhibited higher $SNR_{out}$ values as $SNR_{in}$ increased. Specifically, for CB-CNN and U-Net (i.e., architectures 1 and 2), the median $SNR_{out}$ values increased slightly with $SNR_{in}$. In contrast, the BB-CNN and MWCNN (i.e., architectures 3 and 4) showed a more pronounced linear increase, with slopes of 1.03 and 1.23, and intercepts of 21.72 and 28.89, respectively. The values and statistics of the $SNR_{out}$ for the signals of -20 dB, -15 dB, -10 dB, and -5 dB $SNR_{in}$ are detailed in Table \ref{table:SNR_out_in_archs}.

\begin{table}[ht]
  \caption{$SNR_{out}$ for different architectures.}
  \centering
  \setlength{\tabcolsep}{0.7pt}
  \small
  \begin{tabular}{|ll|rrrr|}
    \hline
    $SNR_{in}$ & $SNR_{out}$ & \makecell[tc]{\scriptsize Architecture 1\\ \scriptsize{CB-CNN}}  & \makecell[tc]{\scriptsize Architecture 2\\ \scriptsize{U-Net}} &  \makecell[tc]{\scriptsize Architecture 3\\ \scriptsize{BB-CNN}} & 
    \makecell[tc]{\scriptsize Architecture 4\\ \scriptsize{MWCNN}} \\
    \hline
    -20 & Median & 4.0805 & 0.5941 & 1.4630 & \textbf{4.3453} \\
        & Mean & 4.1312 & 0.6164 & 1.8801 & \textbf{4.5229} \\
        & Std & 2.1799 & \textbf{1.2463} & 2.8901 & 2.5447 \\
        & CV & \textbf{0.5277} & 2.0217 & 1.5372 & 0.5626 \\
    \hline
    -15 & Median & 8.2749 & 3.5493 & 5.7657 & \textbf{9.8505} \\
        & Mean & 8.1585 & 3.4441 & 6.1535 & \textbf{9.9874} \\
        & Std & 1.8540 & \textbf{1.3122} & 3.3583 & 2.8692 \\
        & CV & \textbf{0.2273} & 0.3810 & 0.5457 & 0.2873 \\
        \hline
    -10 & Median & 10.6791 & 6.4402 & 11.3692 & \textbf{16.6368} \\
        & Mean & 10.7892 & 6.3262 & 11.5154 & \textbf{16.6953} \\
        & Std & \textbf{1.3276} & 1.6023 & 4.0695 & 3.0409 \\
        & CV & \textbf{0.1230} & 0.2533 & 0.3534 & 0.1821 \\
        \hline
    -5 & Median & 11.7160 & 12.0387 & 16.1944 & \textbf{22.8428} \\
       & Mean & 11.8327 & 12.0834 & 16.0205 & \textbf{22.8610} \\
       & Std & \textbf{1.3002} & 2.1512 & 3.8737 & 3.0796 \\
       & CV & \textbf{0.1099} & 0.1780 & 0.2418 & 0.1347 \\
    \hline
  \end{tabular}
  \label{table:SNR_out_in_archs}
\end{table}

Regarding dispersion of performance metrics, the CB-CNN architecture demonstrated the best index, with CV values below 0.53. The CV decreased with SNR improvement, ranging from 0.53 at an input SNR of -20 dB to 0.11 at -5 dB. This stability in output with reduced variation may be attributed to the subsequent convolutional blocks, which likely contribute to noise reduction and elimination of irrelevant variations. It could explain its second-best performance in terms of $SNR_{out}$ for signals with -15 dB or lower. Originally designed for ECG signals with a higher $SNR_{in}$ of -7 dB, this architecture in \cite{DLECG} was trained on signals with consistent amplitudes ranging from 1 mV to 3 mV, simplifying the training process. In contrast, the input data of this study exhibits amplitude variations ranging from 1 mV to 150 mV, potentially contributing to the challenge in denoising signals with a wider voltage range. The drawback of this architecture is encountering difficulties in accurately predicting the signal transitions.

The U-Net architecture shows a reduced uncertainty, as measured using standard deviation, in the output SNR. Nevertheless, the average output SNR is lower than that obtained from the CB-CNN (architecture 1), as shown in Table \ref{table:SNR_out_in_archs}. This observation might be related to the lower performance in estimating the baseline of the signal. This behaviour explains the increased RMSE, MSE, and MAE error metrics, as illustrated by architecture 2 in Fig. \ref{fig:all_metrics_general}. This phenomenon is particularly evident in signals with extended baselines, thus, signals with shorter duty cycles tend to exhibit the lowest SNR improvement.

Overall, the MWCNN architecture demonstrated the highest median output SNR reaching 4.35 dB for input SNR of -20 dB, and 22.84 dB for input SNR of -5 dB. Additionally, across all input SNR values, this architecture exhibited the second-best dispersion based on the CV, surpassed only by the CB-CNN architecture.

\begin{figure}[ht!]
    \centering
    \includegraphics[width=0.9\linewidth]{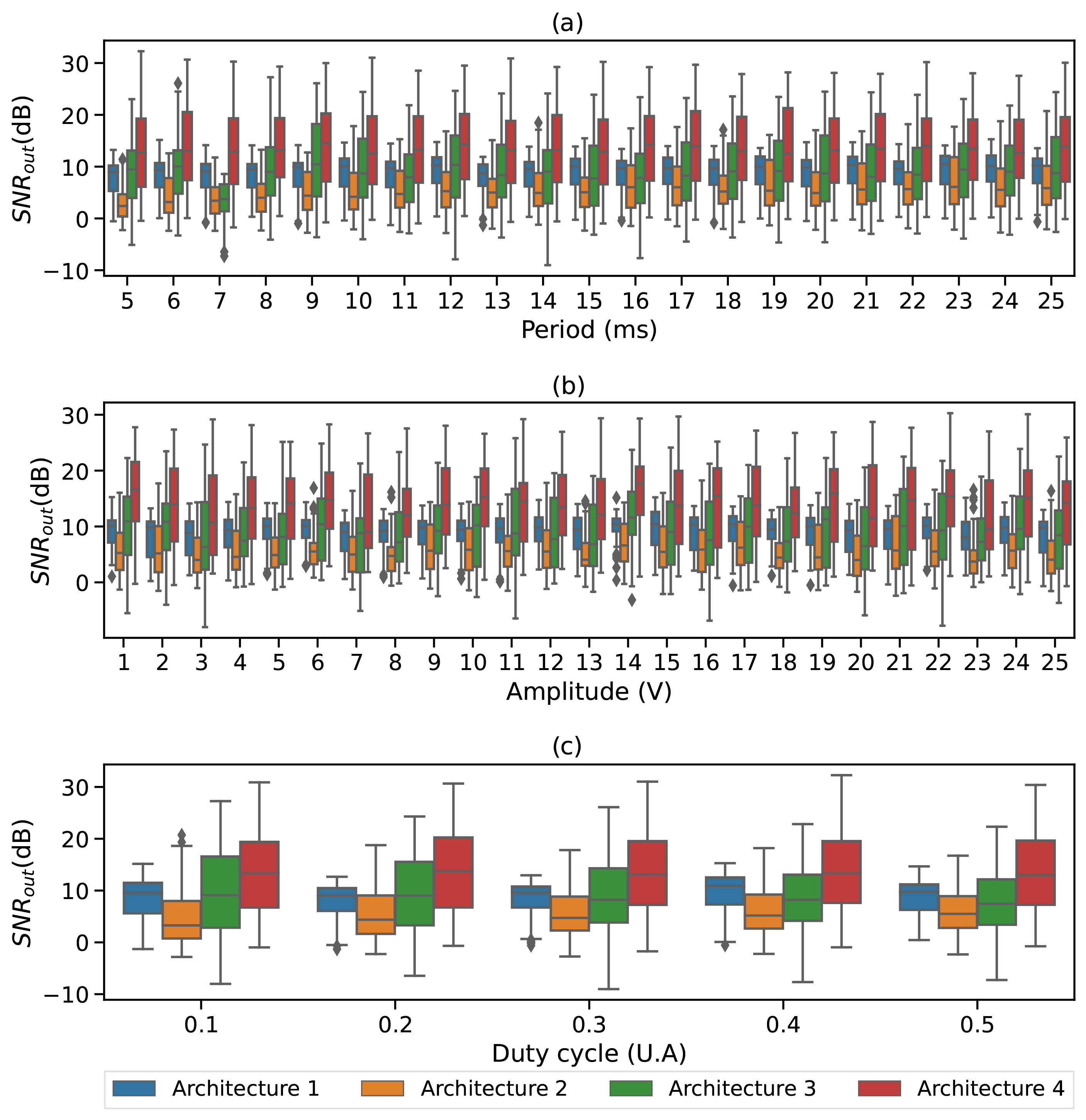}
    \caption[$SNR_{out}$ distributions across signal parameters.]{$SNR_{out}$ distributions across signal parameters. The distributions are shown at each level of (a) period, (b) amplitude up to 25 mV, and (c) duty cycle, for the four architectures.}
    \label{fig:SNR_parameters_all}
\end{figure}

To understand the sources of variability in the output of the architectures, the output SNR was computed across signal parameters, as shown in Fig. \ref{fig:SNR_parameters_all}. 

This figure illustrates consistent performance across signal period, amplitude, and duty cycle. Although amplitudes from 1 to 25 mV are shown for clarity, this behaviour extends to higher amplitudes up to 150 mV. Notably, all architectures maintained consistency against varied parameters, including amplitude, owing to the beneficial effects of Z-score normalisation. However, they encountered challenges in waveform precision, often exhibiting a gradual decline instead of a sharp return to baseline. Addressing the highlighted limitations in future research may involve enhancing signal discrimination. A similar architecture to the MWCNN, utilising the Discrete Wavelet Transform and its inverse for downsampling and upsampling, could be explored in 2-dimensional data. This approach may offer valuable contributions to the field of label-free measurements detection in noisy environments.

\section{Conclusions}
This study evaluated the efficacy of four distinct CNN architectures in denoising synthetic square wave signals, with a bandwidth of 10 kHz, across a range of SNR levels, extending down to -20 dB. Among these architectures, architecture 4, inspired by the MultiWavelet Convolutional Neural Network, emerged as the most effective, achieving a median SNR improvement of 25.9 dB across SNR levels ranging from -5 to -20 dB. Notably, when analysed for $SNR_{in}$ of -5 dB, -10 dB, -15 dB, and -20 dB, this architecture consistently yielded a median SNR improvement ranging from 24.3 dB to 27.8 dB, indicative of its capability to extract signals from noisy backgrounds. Although the performance in terms of $SNR_{out}$ shows a decrease of 1.23 dB for every 1 dB drop in $SNR_{in}$, this model still delivers substantial improvements in noisy conditions and remains highly effective overall.

When we inspect the performance of the four architectures considering signal parameters such as amplitude, period and duty cycle, the architectures exhibited robustness.
The consistency of $SNR_{out}$ was demonstrated across varying signal parameters, including period (5-25 ms), duty cycle (0.1-0.5), and amplitude (1-150 mV). It is important to note that performing a pre-processing stage to standardise the input signal, following the Z-score normalisation described in section \ref{Zscore}, is essential to improve the performance on low-amplitude signals, thus enhancing the generalisability of the denoising approach.

Comparing the variability of different architectures, the CB-CNN and the U-Net, namely architectures 1 and 2, demonstrated lower standard deviation of all the metrics, such as output SNR and RMSE. In particular, the CB-CNN consistently exhibited greater homogeneity across all the input SNR studied, with CV values lower than 0.53, indicating more stable output characteristics. Conversely, the BB-CNN architecture and the MWCNN (architectures 3 and 4) exhibited increased variability in performance metrics such as output SNR. Although, architecture 4 showcased the highest median $SNR_{out}$ performance, it encountered challenges in maintaining homogeneity comparable to the CB-CNN architecture. 

The identification of the baseline of the signal is generally well established. However, the remaining challenges primarily involve enhancing the precision of shape retrieval. Specifically, the architecture 4, multilevel wavelet CNN structure, demonstrated superior performance to the rest of the architectures for all input SNR levels. Therefore, it might be worth investigating its mechanism for reconstructing the signal in comparison to the other architectures.
Furthermore, future studies could investigate the effectiveness of custom loss function on shape retrievals in comparison MSE that was used in this study.

In summary, the comparison of different CNN architectures in this study emphasises the suitability of CNNs, particularly the MWCNN-inspired architecture, for denoising low amplitude signals with a 10 kHz bandwidth, including variations in period, duty cycle, and amplitude. These findings hold significant promise for the application of CNN-based denoising techniques to extract signals from extremely noisy measurement conditions which could benefit new approaches to measuring bioelectrical signals optically.

\section*{Acknowledgment}
Funding: This work was supported by the Engineering and Physical Sciences Research Council [grant number EP/X018024/1]; A.F.Q. acknowledges support from the ANID National Agency for Research and Development through the Master's Scholarship Abroad Programme, Becas Chile, Call 



 \bibliographystyle{elsarticle-num} 
 \bibliography{dl-cnn-bib}





\end{document}